\newif\iffigs\figstrue

\documentclass[paper, 12pt, letterpaper]{JHEP}

\def\Bbb{\bf}
\def\C{{\Bbb C}}

\def\Z{{\Bbb Z}}

\def\bearray{\begin{eqnarray}}
\def\eearray{\end{eqnarray}}
\def\bearraynn{\begin{eqnarray*}}
\def\eearraynn{\end{eqnarray*}}
\def\bfig{\begin{figure}}
\def\efig{\end{figure}}

\def\opeq#1{\advance\lineskip#1 \advance\baselineskip#1
        \advance\lineskiplimit#1}

\def\boldone{\relax{\rm 1\kern-.35em 1}}

\newtheorem{Proposition}{Proposition}[section]

\newtheorem{Theorem}{Theorem}[section]
\newtheorem{Lemma}{Lemma}[section]
\newtheorem{Corrolary}{Corrolary}[section]

\newcommand{\be}{\begin{equation}}
\newcommand{\ee}{\end{equation}}
\newcommand{\bea}{\begin{eqnarray}}
\newcommand{\eea}{\end{eqnarray}}

\newcommand{\bp}{\begin{Proposition}}
\newcommand{\ep}{\end{Proposition}}
\newcommand{\bt}{\begin{Theorem}}
\newcommand{\et}{\end{Theorem}}
\newcommand{\bl}{\begin{Lemma}}
\newcommand{\el}{\end{Lemma}}
\newcommand{\bc}{\begin{Corrolary}}
\newcommand{\ec}{\end{Corrolary}}
\newcommand{\nn}{\nonumber}

\usepackage{graphics}

\title{Generalized complexes and string field theory}

\author{C.~I.~Lazaroiu\\C.~N.~Yang Institute for Theoretical Physics\\
SUNY at Stony BrookNY11794-3840, U.S.A.\\calin@insti.physics.sunysb.edu} 

\abstract{I discuss the axiomatic framework of (tree-level) associative 
open string field 
theory in the presence of D-branes by considering the natural extension of the 
case of a single boundary sector. This leads to a formulation which is 
intimately connected with the mathematical theory of differential graded categories.
I point out that a generic string field theory as formulated within this framework
is not closed under formation of D-brane composites 
and as such does not allow for a unitary description of D-brane dynamics. This implies 
that the collection of boundary sectors of a generic string field theory with D-branes 
must be extended by inclusion of all possible D-brane composites. 
I give a precise formulation of a weak unitarity constraint and 
show that a minimal extension which is unitary in this sense can always be obtained 
by promoting the original D-brane category to an enlarged category constructed 
by using certain generalized complexes of D-branes. I give a detailed construction 
of this extension and prove its closure under formation of D-brane composites. 
These results amount to a completely general description of D-brane composite formation 
within the framework of associative string field theory. }

\begin{document}

\tableofcontents

\pagebreak

\section{Introduction}

Recent work on D-brane physics has centered around the problem of describing 
the most general D-branes which can be introduced in a given closed string background.
In the language of \cite{top}, this can formulated as follows:

{\bf Problem} Given a closed string background, classify all of its open-closed 
extensions (i.e. describe all open-closed string theories whose closed string sector 
coincides with the given bulk theory).

Here D-branes are understood in an abstract sense, namely as 
{\em boundary sectors of a string theory}. 
This definition does not distinguish 
`composite' objects such as stable non-BPS D-branes from those which can be 
described semiclassically 
through boundary conditions in a nonlinear sigma model (`bootstrap').

Although not explicitly formulated as such, this problem forms the core 
of efforts to identify the correct classification of 
D-brane charges \cite{Witten_K,K_bos,Moore_top}
and underlies recent work
\cite{Douglas_Kontsevich} aimed at understanding the physical foundations 
of the derived category constructions 
which enter the homological mirror symmetry conjecture \cite{Kontsevich}.

The main difficulty of this approach comes from the observation 
that a consistent string theory with D-branes 
must include a dynamically complete set of boundary 
sectors. This is essentially a unitarity constraint, which can be understood 
as follows. Consider a string theory whose boundary sectors are 
described by standard D-branes, i.e. through imposing Dirichlet boundary 
conditions on certain submanifolds of the space-time manifold (we assume 
for the moment 
that the underlying conformal field theory admits a sigma model 
interpretation). It is by now well-understood that a generic
\footnote{BPS configurations in type II theories, which were 
historically studied most intensively, are highly non-generic (they form 
a set of `measure zero' in the class of all conceivable D-brane 
configurations).}
D-brane configuration will be unstable, a phenomenon which has been 
studied for parallel D-branes in both superstring field theory and bosonic 
string field theory (see \cite{tachyon} for a very partial list of references). 
In its standard realization, this 
follows from the existence of tachyon modes which condense in some nontrivial 
fashion. Since condensation phenomena involve off-shell 
string dynamics, the end result of such a process cannot be generally 
described in sigma model language (i.e. as a standard D-brane, defined 
through boundary conditions on the worldsheet fields). This basic fact implies 
that a unitary description of the string theory requires the inclusion 
of all such condensation products (which we shall 
call `generalized 
D-branes') along with the original D-branes described through boundary 
conditions. Indeed, a minimal requirement for unitary dynamics is that the 
the set of boundary sectors be closed under all physical processes. 

This observation has far reaching implications for the general structure 
of open string field theories.  
In this paper, I shall focus on its 
consequences for {\em associative} tree level 
open string theories in the presence of D-branes. More precisely,  
I shall 
argue that systematic consideration of condensation processes leads to a 
description of D-brane composites through a certain type of {\em generalized complexes},
which (as I shall show somewhere else) 
turn out to be related to the twisted 
complexes of \cite{Bondal_Kapranov} in the particular 
case of topological string theories. Generalized complexes form a category 
which extends the D-brane category of the given theory.

Since category theory is not a traditional subject of interest among 
physicists, let me give a short explanation of its relevance. 
Categories arise  
in oriented open string theories in the presence of D-branes via the simple 
observation that one can think of D-branes as objects $a$ and of the states 
of open strings stretched between them as morphisms defining string state 
spaces $Hom(a,b)$. 
In this case, the 
natural composition of morphisms is given by the double (or basic) 
string product
$r^{(2)}:Hom(b,c)\times Hom(a,b) \rightarrow Hom(a,c)$, which is related 
to the triple string correlator on the disk via:
\be
\langle u,v,w\rangle =(u, r_2(v,w))
\ee
for all $u\in Hom(c,a), v\in Hom(b,c), w\in Hom(a,b)$. In this relation, 
$(.,.)$ is the double {\em conformal} correlator on the disk. In conformal field 
theory language, the elements of $Hom(a,b)$ correspond to boundary condition 
changing operators if $a\neq b$, while for $a=b$ we have usual boundary 
operators associated with the D-brane $a$, which give the associated space of 
endomorphisms, $End(a)$. To obtain a category in the standard mathematical 
sense, the composition of morphisms must be associative\footnote{One must also 
have identity operators $Id_a\in Hom(a,a)$ for all $a$, which are provided 
by the boundary vacua $|0_a\rangle$\cite{top}.}. This can always be 
assured by going on-shell, i.e. considering the states in $Hom(a,b)$ only up 
to BRST exact states. It is well-known, however, that an on-shell analysis 
does  not suffice to describe string theory dynamics (in the same way that it 
does not suffice to describe the dynamics of a gauge theory). In particular, 
the on-shell formalism is not appropriate when studying formation of D-brane 
composites. Therefore, 
taking BRST cohomology is not advisable as an intermediate step, and should 
only be done at the end of the analysis of any given problem. In physical 
terms, this means that we should approach nontrivial dynamical issues 
with the tools of string field theory.
It follows that 
one should take $Hom(a,b)$ as being {\em off-shell} string state spaces. 
In this case, one encounters a potential problem, since it is 
well-known that the off-shell product $r^{(2)}$ need not be associative. 
In general, $r^{(2)}$ is only associative up to homotopy \cite{Gaberdiel}, 
i.e. up to BRST exact terms, which involve the BRST variation of the triple 
product.  This forces one to the conclusion that a general off-shell analysis 
should involve a certain generalization of the classical notion of category, 
which was considered in \cite{Fukaya, Fukaya2} and is known as an {\em $A_\infty$} 
category. The general mathematical theory of $A_\infty$ categories is still 
in its infancy (see, however, \cite{Keller}) 
and will be avoided in this paper. Instead, I shall restrict 
to the case when the product $r^{(2)}$ happens to be associative 
{\em off-shell}, the so-called case of {\em associative open string theory}.
Well-known examples of such theories are the bosonic string field theory 
of \cite{Witten_SFT} 
and the (open) holomorphic Chern-Simons theory 
of \cite{Witten_CS} which describes the dynamics of the open topological 
B string. 

To gain a general understanding of associative string field 
dynamics one must first enlarge the axiomatic framework of \cite{Witten_SFT}
to allow for the presence of D-branes. As suggested above, the correct 
formulation in the associative case naturally involves category theory. 
Since the BRST operator acts as a derivation of the string product, one 
is lead to consider so-called {\em differential graded categories}, 
i.e. categories whose morphism spaces $Hom(a,b)$ are graded vector spaces 
carrying the structure of complexes, and whose differentials 
are compatible with morphism 
compositions. The relevant axioms, which are a straightforward extension of 
those considered in \cite{Witten_SFT}, are discussed in Section 3 below. 

It is thus clear that 
basic `structural' issues in D-brane physics (such as the classification 
of D-branes and their charges, or the extra structure which describes 
`unitary' brane dynamics) should be approached with category-theoretic tools.
In this paper, I follow this point of view by providing an analysis of D-brane 
composite formation and of the unitarity constraint. This analysis is rather abstract, 
a price we have to pay for its extreme generality, but I hope that it successfully 
addresses some basic aspects of the structure of open string theory with 
D-branes. It also raises a set of tantalizing questions which will be briefly 
mentioned in the conclusions.

A complete study of the realization of our constructions in 
physically interesting string theories can be a daunting task, given the 
currently incomplete 
understanding of their basic algebraic and analytic aspects (not to mention  
serious computational difficulties). 
For example, very little 
is understood about the correct completion of the bosonic open string algebra 
of \cite{Witten_SFT}. For the superstring, even more basic structural issues 
are as yet unclarified. For these reasons, a detailed 
analysis is currently 
limited to `toy models' such as the string field theories of topological 
strings. It is natural to expect such an analysis to make contact 
with the mathematical program of homological mirror symmetry 
\cite{Kontsevich, Polishchuk, Polishchuk2, 
Barannikov_formality, Barannikov1, Barannikov2, Kontsevich_recent, Zaslow_hom, 
Fukaya, Fukaya2, Fukaya_trees, ST, Thomas} (see also 
\cite{Vafa, Vafa_mirror, Hori, Kapustin, SYZ, LYZ, Hofman, boundary})
and we shall make some connections in a companion paper. 

The organization of this paper is as follows. 
In section 2, we discuss the axiomatic 
framework of associative open string field theory in the presence of D-branes, by 
formulating it in the language of differential graded categories. In Section 3, we 
recall the standard construction of the string field theory moduli space and 
relate it to modern deformation theory. In Section 4, we consider D-brane 
composite 
formation in a string field theory with D-branes, by approaching it with the tools
of Section 3. We show that shifts of the string vacuum allow for a systematic 
description of D-brane condensation and give a precise formulation of the 
weak unitarity (`quasiunitarity') 
condition that a given open string theory is closed under D-brane formation.
This translates the unitarity constraint into a condition on the underlying 
differential graded category. Section 4 shows how one can implement this 
constraint by starting with an arbitrary open string field theory with D-branes 
and considering a certain extension of its underlying category, which we 
call its `quasiunitary cover'. We show that this extension can be built 
as a category of `generalized complexes' of objects of the original category, 
and prove that it satisfies the quasiunitarity constraint. Section 5 suggests 
some directions for further research. The appendix collects some basic concepts 
relevant to differential graded categories.

\section{The axiomatic framework of associative
open string field theory with D-branes}

It can be shown  
that the a general tree level open string field theory 
in the presence of D-branes can be described through an $A_\infty$ category
\cite{Fukaya} (this follows from a relatively straightforward, but rather 
tenuous, extension of the axiomatic analysis of 
\cite{Zwiebach_open, Gaberdiel}). In this paper I shall restrict to the 
associative case, which can be described in a more 
standard mathematical language as follows
\footnote{Note that we are using conventions 
which are slightly different from those of \cite{Gaberdiel}, but better adapted to making 
contact with the physics literature. For the bosonic string, our grading $|.|$ 
is given by the ghost number $gh$, while the grading of \cite{Gaberdiel} 
is given by 
$1-gh$. Our conventions also differ from those of \cite{Witten_SFT}.}: 

\

A (tree level) open string field theory in the presence of D-branes 
is given by the following data: 

\

(1) A differential graded $\C$-linear category ${\cal A}$

\

(2) For each pair of objects $a, b$ of ${\cal A}$, an invariant 
nondegenerate bilinear and graded-symmetric form 
$_{ab}(.,.)_{ba}:Hom(a,b)\times Hom(b,a)\rightarrow \C$ of 
degree $3$.

\

Let us explain each piece of data.
By a 
{\em $\C$-linear category} we mean a category whose morphism spaces are 
complex vector spaces and whose morphism compositions are bilinear maps. 
A {\em graded linear category} is a linear category whose morphism spaces 
are $\Z$-graded, i.e. $Hom(a,b)=\oplus_{k\in \Z}{Hom^k(a,b)}$, and 
such that morphism compositions are homogeneous of degree zero, i.e.: 
\be
|uv|=|u|+|v|~~{\rm~for~all~homogeneous~}u\in Hom(b,c),~v\in Hom(a,b)~~,
\ee 
where $|.|$ denotes the degree of a homogeneous element.
In a {\em differential graded linear category} (dG category, for short), the morphism spaces 
$Hom(a,b)$ are further endowed with nilpotent operators $Q_{ab}$ 
(i.e. $Q_{ab}^2=0$) 
of degree $+1$ which act as derivations of morphism compositions:
\be
\label{der}
Q_{ac}(uv)=Q_{bc}(u)v+(-1)^{|u|}uQ_{ab}{v}~{\rm~for~homogeneous~}u\in Hom(b,c),~
v\in Hom(a,b)~~.
\ee
This can also be expressed as the requirement that the composition 
maps 
$Hom(b,c)\otimes Hom(a,b)\rightarrow Hom(a,c)$ are morphisms of complexes, 
where $Hom(b,c)\otimes Hom(a,b)$ carries the structure of total complex
induced from its components, i.e.:
\bea
\label{tc}
[Hom(b,c)\otimes Hom(a,c)]^k&=&\oplus_{i+j=k}{Hom^i(b,c)\otimes Hom^j(a,b)}~~\nn\\
Q(u\otimes v)&=&Q_{bc}u\otimes v+ (-1)^{|u|}u\otimes    Q_{ab}v~~.
\eea
Graded symmetry of the bilinear forms means: 
\be
_{ab}(u,v)_{ba}=(-1)^{|v||u|}_{ba}(v,u)_{ab}
\ee 
for homogeneous elements $u\in Hom(a,b)$ and $v\in Hom(b,a)$. 
The degree $3$ constraint is:
\be
_{ab}(u,v)_{ba}=0~~{\rm~unless~}|u|+|v|=3~~.
\ee
The bilinear form is required to be invariant with respect 
to the differentials $Q_{ab}$ and with respect to morphism compositions:
\be
\label{d_invariance}
_{ab}(Q_{ab}(u),v)_{ba}+(-1)^{|u|}(u,Q_{ba}(v))=0~~{\rm~for~}
u\in Hom(a,b),~v\in Hom(b,a)~~
\ee
and
\be
_{ca}(u,vw)_{ac}=_{ba}(uv,w)_{ab}~~{\rm~for~}u\in Hom(c,a),~v\in Hom(b,c),~
w\in Hom(a,b)~~.
\ee

In physical theories such as the bosonic string, one also has antilinear involutions 
obeying certain constraints involving the bilinear forms and the BRST operators. This 
extra structure need not be present in a topological field theory\footnote{For example, 
it is irrelevant for the B-model
\cite{Witten_NLSM, Witten_mirror, Witten_CS}, whose string field action is 
allowed to be complex.}
and will not be considered in this paper. It is not hard to extend the discussion below 
by including such conjugations, but this adds little to our results while cluttering 
the presentation.

The physical interpretation of this data is as follows. The objects $a$ 
are identified as D-branes (in an abstract sense), while 
$Hom(a,b)$ is the (off-shell) state space of 
strings stretching from $a$ to $b$. These will be called the 
{\em boundary sectors} of the theory. The sectors $Hom(a,a)=End(a)$ are the 
diagonal boundary sectors, while $Hom(a,b)$ with $a\neq b$ are the 
off-diagonal, or boundary condition changing sectors. 
The nilpotent operators $Q_{ab}$  give the BRST charge in the sectors 
$Hom(a,b)$, while the $\Z$-grading of $Hom(a,b)$ is induced by 
a grading on the boundary worldsheet fields. 
In a physical theory such as the open bosonic string, this is simply the 
ghost number degree, while for A/B topological strings it is the grading 
associated with the anomalous $U(1)$ symmetry of the topological (`twisted') 
$N=2$ superconformal algebra.

The nondegenerate bilinear forms $_{ab}(.,.)_{ba}$  induce  linear 
isomorphisms:
\be
bpz_{ab}:Hom(a,b)\stackrel{\approx}{\longrightarrow}Hom(b,a)^*~~
\ee
through the standard prescription:
\be
bpz_{ab}(u)(v):=_{ba}(v,u)_{ab}~~{\rm~for~all~}u\in Hom(a,b),~v\in Hom(b,a)~~.
\ee
These are the usual BPZ conjugations.

\subsection{On-shell formalism}

One can recover the on-shell description by taking the BRST cohomology of all 
boundary state spaces $Hom(a,b)$. It is easy to see that this produces 
a graded category, the so-called {\em cohomology category} 
of the dG category ${\cal A}$( see Appendix 1). 
This has the same objects as ${\cal A}$ 
but morphism spaces given by the BRST cohomology:
\be
Hom_{H({\cal A})}(a,b)=H_Q^*(Hom_{\cal A}(a,b))~~.
\ee

The morphism compositions of $H({\cal A})$ are 
induced from those of ${\cal A}$ and are well-defined by virtue of condition 
(\ref{der}). Moreover, condition (\ref{d_invariance}) implies that 
the bilinear forms of ${\cal A}$ descend to forms $_{a,b}(.,.)_{ba}$ 
on $Hom_{H({\cal A})}(a,b)\otimes Hom_{H({\cal A})}(b,a)$, which are 
invariant with respect to composition of morphisms in $H({\cal A})$. 
This recovers the boundary part of the data discussed in \cite{top}.

\subsection{The total boundary state space}

The structure above can be related to the more familiar data of \cite{Witten_SFT}
(see \cite{Thorn} for a review) 
by defining the total boundary state space ${\cal H}=\oplus_{a,b}{Hom(a,b)}$.
Then one defines the total BRST charge via:
\be
Q=\oplus_{a,b}{Q_{ab}}~~,
\ee
and the total boundary composition 
${\cal H}\times {\cal H}\rightarrow {\cal H}$:
\be
uv=\oplus_{ac}{\left[\sum_{b}{u_{bc}v_{ab}}\right]}~~,
\ee
for $u=\oplus_{ab}{u_{ab}}$, $v=\oplus_{ab}{v_{ab}}$ with 
$u_{ab},v_{ab}\in Hom(a,b)$. Moreover, the bilinear forms $_{ab}(.,.)_{ba}$ 
induce the total boundary form:
\be
(u,v)=\sum_{ab}{_{ab}(u_{ab},v_{ba})_{ba}}~~.
\ee
It is easy to see that the resulting `total' data satisfies the axioms 
of \cite{Witten_CS}, i.e.: 

(I) ${\cal H}$ together with $Q$ and the total boundary composition is a 
differential graded associative algebra

(II) The total boundary bilinear form $(.,.)$ is nondegenerate and invariant 
with respect to the boundary BRST operator and the boundary total product. 

However, our theory contains extra structure, encoded in the 
category-theoretic 
properties of $Q$, $(.,.)$ and the boundary product. 
This structure amounts to the statement that the total boundary data 
decompose in the form given above. For example, the boundary product 
must be nonzero only on subspaces of the form 
$Hom(b,c)\times Hom(a,b)$, which it must take into the subspace $Hom(a,c)$ of 
${\cal H}$. Also note that, in practice, the number of objects (D-branes) 
will  typically be infinite (infinite of the power of the continuum, 
which makes the interpretation in terms of the 
total boundary state space somewhat tenuous).

The string field action on the total boundary space has the form considered in 
\cite{Witten_CS} (we assume that the given background satisfies the 
string equations of motion, so that the action does not contain a linear term):
\be
\label{action}
S(\phi)=\frac{1}{2}(\phi,Q\phi)+\frac{1}{3}(\phi, \phi\phi)~, 
\ee
where the string field $\phi=\oplus_{a,b}{\phi_{ab}}$ (with 
$\phi_{ab}\in Hom(a,b)$) is constrained to be of degree one
\footnote{In a theory admitting antilinear involutions , 
one also imposes a reality constraint on $\phi$.}:
\be\
|\phi_{ab}|= 1~~.
\ee 
The string field action can be trivially expanded in boundary sectors:
\be
S(\phi)=\frac{1}{2}{\sum_{ab}{_{ab}(\phi_{ab},Q_{ba}\phi_{ba})_{ba}}}+
\frac{1}{3}{\sum_{abc}{_{ca}(\phi_{ca},\phi_{bc}\phi_{ab})_{ac}}}~~.
\ee 
In a realistic case (backgrounds containing an infinity of D-branes), 
this is of course a formal expression, unless one assumes that only a finite 
number of boundary sectors can be simultaneously excited (in this case, 
the string field is restricted to have nonzero value only in a finite number of 
boundary sectors).

\section{The string field theory moduli space}

In this section we recall the basic procedure for building the string field theory 
moduli space and discuss its connection with shifts of the string vacuum. All of the results 
presented below are well-known or implicit in the literature, and my only contribution 
is to reformulate some of them in the language of deformation theory.

Let us consider a string field theory based on a dG algebra ${\cal H}$, which 
can be the boundary state space of a theory with one D-brane or the total boundary 
space of a theory with multiple D-branes. In this section, we work at the level 
of the algebra ${\cal H}$, temporarily ignoring the category structure.

Given such a theory, a shift of  
the background (i.e. of the associated string vacuum) corresponds 
to a translation $\phi\rightarrow \phi+q$ 
of the string field,  which induces a translation 
$S(\phi)\rightarrow S'(\phi)=S(\phi+q)$ of the string field 
action\footnote{That is, we set $\phi=\phi'+q$ so that $S(\phi)=S(\phi'+q)=
S'(\phi')$, then we rename $\phi'$ as $\phi$.}. Since the string field has 
degree one, we must assume $|q|=1$.
The shifted action can be expanded as:
\be
\label{shifted_action}
S'(\phi)=S(\phi+q)=S(q)+(r_0,\phi)+{\tilde S}(\phi)~~,
\ee
where:
\bea
r_0&=&Qq+q^2~~\\
{\tilde S}(\phi)
&=&\frac{1}{2}(\phi, (Q+2q)\phi)+\frac{1}{3}(\phi, \phi\phi)~~.
\eea
One can bring ${\tilde S}$ 
to the form (\ref{action}), by defining a shifted BRST 
operator:
\be
\label{Q'}
Q'u=Qu+[q,u]~~ ({\rm~for~any~}u\in {\cal H}~)~~,
\ee
where $[,]$ is the graded commutator on ${\cal H}$ induced by the  
associative product :
\be
[u,v]=uv-(-1)^{|u||v|}vu~~.
\ee
Then $Q'\phi=(Q+2q)\phi$ (since $|q|=|\phi|=1$) and we can write:
\be
{\tilde S}(\phi)=
\frac{1}{2}(\phi, Q'\phi)+\frac{1}{3}(\phi, \phi\phi)~~.
\ee
The linear term in (\ref{shifted_action}) signals the 
presence of string tadpoles in an expansion around the new vacuum. 
The condition that this vacuum satisfies the string equations of motion 
(i.e. the tadpole cancellation constraint) takes the form $r_0=0$, 
i.e. 
\be
\label{cons}
Qq+qq=0~~.
\ee
On the other hand, it is easy to compute:
\be
(Q')^2u=[Qq+q^2, u]~~,
\ee
so that the operator  $Q'$ is nilpotent on ${\cal H}$ 
precisely when the tadpole 
cancellation condition (\ref{cons}) holds. In this case, $Q'$ is obviously 
a degree one derivation of the associative product, so that 
$({\cal H}, Q',\cdot)$ is a dG algebra. 
Since the constant term $S(q)$ is irrelevant, it 
follows that a shift of the vacuum is equivalent with the replacement 
$S(\phi)\rightarrow {\tilde S}(\phi)$, which in turn amounts to 
the modification $Q\rightarrow Q'$ of the BRST charge. Moreover, the tadpole
cancellation constraint (\ref{cons}) is equivalent with the condition that 
the deformed structure $({\cal H}, Q',\cdot)$ remains a dG algebra.
 
One can recognize the standard ingredients of deformation theory 
\cite{Kontsevich_def_q}(see also \cite{Manin}). 
Indeed, since $|q|=1$, we have $[q,q]=2q^2$ 
and we can re-write the basic constraint (\ref{cons}) in the standard form:
\be
\label{qmc}
Qq+\frac{1}{2}[q,q]=0~~.
\ee
This is the Maurer-Cartan equation in the dG Lie algebra 
$({\cal H}, d, [.,.])$ induced by our differential graded associative algebra. 
Moreover, two infinitesimal shifts $q$, $q'$
should be identified if they are related through:
\be
q'=q+\alpha~~,
\ee
where $\alpha$ is of the form:
\be
\label{gauge}
\alpha=Q'\beta=Q\beta+[q,\beta]~~,
\ee
with $\beta$ a degree zero element of ${\cal H}$. 
It follows the moduli space of 
string vacua is the moduli space associated with the deformation problem 
described by the Maurer-Cartan equation (\ref{qmc}). Its tangent space at the origin 
(i.e. at the point corresponding to the original vacuum) is given by the linearization 
of (\ref{qmc}) and the zeroth order approximation to (\ref{gauge}), 
which define the first degree $Q$-cohomology $H_Q^1({\cal H})$.

\section{Shifts of a string vacuum with D-branes and formation of D-brane composites}

In the case of multiple D-branes, one can apply the general discussion of the previous 
subsection to the dG associative algebra ${\cal H}=\oplus_{a,b}Hom(a,b)$. In this 
case, a deformation $q$ has the expansion $q=\sum_{a,b}{q_{ab}}$, with $q_{ab}
\in Hom^1(a,b)$. 
Expanding the equations above into boundary 
sectors leads to the tadpole cancellation constraint:
\be
\label{tadpole_expanded}
Q_{ab}q_{ab}+\sum_{c}{q_{cb}q_{ac}}=0~~,
\ee
and to the following expression for the shifted BRST charge:
\be
\label{Q'_expanded}
Q'u=\oplus_{a,b}{\left[
Q_{ab}u+\sum_{c}{\left(q_{cb}u_{ac}-(-1)^{|u_{cb}|}u_{cb}q_{ac}\right)}
\right]}~~, 
\ee
where $u=\oplus_{ab}{u_{ab}}$, with $u_{ab}\in Hom(a,b)$. The degree one condition on 
$q_{ab}$ assures that $q$ is a homogeneous element of degree one, and thus $Q'$ is 
a derivation of degree $+1$. It 
also assures that $q$ can be thought of as a shift of the 
degree one string field $\phi$.

\subsection{D-brane composites and pseudocomplexes}

Let us define a {\em pseudocomplex} to be given by a set $S$ of objects of ${\cal A}$ 
together with a set of degree one morphisms $q_{ab}\in Hom^1(a,b)$ ($a,b \in S$) which 
satisfy the tadpole cancellation constraint (\ref{tadpole_expanded}). More precisely, this 
condition has to be satisfied for {\em all} objects $a,b$ in ${\cal A}$, where we 
define $q_{de}=0$ if $d$ or $e$ does not belong to the set $S$. An example of 
pseudocomplex is shown in Figure 1.

\hskip 0.8 in
\begin{center} 
\scalebox{0.6}{\begin{picture}(0,0)%
\includegraphics{pc.pstex}%
\end{picture}%
\setlength{\unitlength}{4144sp}%
\begingroup\makeatletter\ifx\SetFigFont\undefined%
\gdef\SetFigFont#1#2#3#4#5{%
  \reset@font\fontsize{#1}{#2pt}%
  \fontfamily{#3}\fontseries{#4}\fontshape{#5}%
  \selectfont}%
\fi\endgroup%
\begin{picture}(6345,3746)(1081,-2941)
\put(6166,-2581){\makebox(0,0)[lb]{\smash{\SetFigFont{14}{16.8}{\familydefault}{\mddefault}{\updefault}
$q_{ab}$
}}}
\put(5086,-2941){\makebox(0,0)[lb]{\smash{\SetFigFont{14}{16.8}{\familydefault}{\mddefault}{\updefault}
$a$
}}}
\put(6661,-2131){\makebox(0,0)[lb]{\smash{\SetFigFont{14}{16.8}{\familydefault}{\mddefault}{\updefault}
$q_{db}$
}}}
\put(4906,-1546){\makebox(0,0)[lb]{\smash{\SetFigFont{14}{16.8}{\familydefault}{\mddefault}{\updefault}
$q_{ca}$
}}}
\put(5941,-601){\makebox(0,0)[lb]{\smash{\SetFigFont{14}{16.8}{\familydefault}{\mddefault}{\updefault}
$c$
}}}
\put(6841,-1096){\makebox(0,0)[lb]{\smash{\SetFigFont{14}{16.8}{\familydefault}{\mddefault}{\updefault}
$q_{cb}$
}}}
\put(5941,-1366){\makebox(0,0)[lb]{\smash{\SetFigFont{14}{16.8}{\familydefault}{\mddefault}{\updefault}
$q_{cd}$
}}}
\put(5716,-1816){\makebox(0,0)[lb]{\smash{\SetFigFont{14}{16.8}{\familydefault}{\mddefault}{\updefault}
$d$
}}}
\put(7426,-1816){\makebox(0,0)[lb]{\smash{\SetFigFont{14}{16.8}{\familydefault}{\mddefault}{\updefault}
$b$
}}}
\put(3736,-1096){\makebox(0,0)[lb]{\smash{\SetFigFont{14}{16.8}{\familydefault}{\mddefault}{\updefault}
$f$
}}}
\put(1261,-1051){\makebox(0,0)[lb]{\smash{\SetFigFont{14}{16.8}{\familydefault}{\mddefault}{\updefault}
$e$
}}}
\put(3106,-16){\makebox(0,0)[lb]{\smash{\SetFigFont{14}{16.8}{\familydefault}{\mddefault}{\updefault}
$i$
}}}
\put(4231,479){\makebox(0,0)[lb]{\smash{\SetFigFont{14}{16.8}{\familydefault}{\mddefault}{\updefault}
$g$
}}}
\put(1081,524){\makebox(0,0)[lb]{\smash{\SetFigFont{14}{16.8}{\familydefault}{\mddefault}{\updefault}
$h$
}}}
\put(6436,-1591){\makebox(0,0)[lb]{\smash{\SetFigFont{14}{16.8}{\familydefault}{\mddefault}{\updefault}
$q_{bd}$
}}}
\end{picture}
}
\end{center}
\begin{center} 
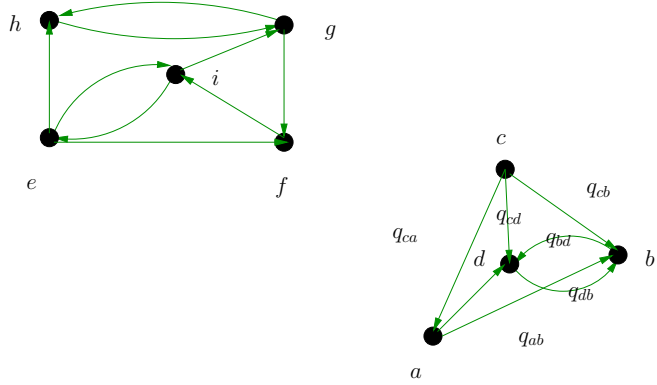
Figure  1. {\footnotesize A pseudocomplex with two connected components. 
For the second component, we also indicate the labeling of  
morphisms. Note that the objects $a..i$ sitting at the nodes are all distinct.}
\end{center}

In view of the above, 
pseudocomplexes describe vacuum shifts of our string theory, and allow one to build its 
vacuum moduli space by solving the moduli problem defined by the Maurer-Cartan equation 
(\ref{tadpole_expanded}). The fact that we have multiple boundary sectors implies that a 
deformation through such a complex corresponds to condensation 
of boundary states $q_{ab}$, 
leading to a D-brane composite. Indeed, the standard interpretation of a shift by $q$ is 
that the components $q_{ab}$ of the string field acquire VEV's, which is the original physical 
motivation for shifting the string vacuum. This suggests that {\em generalized 
complexes describe D-brane composites}. To justify this more formally, notice that the shifted 
theory does not preserve the original decomposition ${\cal H}=\oplus_{a,b}{{\cal H}_{ab}}$ 
into boundary sectors, in the sense that the shifted BRST operator $Q'$ does not map 
every subspace ${\cal H}_{ab}$ into ${\cal H}_{ab}$, thus violating our axioms. This can be remedied 
by taking the approach of \cite{top}, namely we view the data 
$({\cal H}, Q', \cdot)$ as fundamental and look for a new decomposition of ${\cal H}$ into 
boundary sectors such that the axioms of Section 2 are satisfied. 

To make this precise, notice that a pseudocomplex 
can be viewed as a graph $\Gamma$ 
whose nodes are the objects $a\in S$ and whose arrows are given by 
those 
morphisms $q_{ab}$ which are {\em non-vanishing} 
(visualized as an arrow from $a$ to $b$). We say that the complex
$q$ is {\em irreducible} (or {\em connected}) 
if this graph is connected, i.e. any two distinct objects $a,b\in S$
can be connected through a sequence of arrows belonging to the graph.  

\subsubsection{Shifts by irreducible $q$}

For an irreducible $q$, the total 
boundary space ${\cal H}$ admits the decomposition:
{\footnotesize \be
{\cal H}=[\oplus_{a,b\in S}{Hom(a,b)}]\oplus 
[\oplus_{a,b {\not \in} S}{Hom(a,b)}]
\oplus 
\left(\oplus_{b {\not \in} S}{[\oplus_{a\in S}{Hom(a,b)}]}\right)\oplus 
\left(\oplus_{a {\not \in} S}{[\oplus_{b\in S}{Hom(a,b)}]}\right)~~,\nn
\ee}\noindent which preserves all axioms 
if the terms in square brackets are viewed as the new boundary sectors.
This amounts to viewing the shifted vacuum as the category ${\cal A}_q$ 
obtained from ${\cal A}$ by collapsing 
the set of objects $S$ to a single new object $*$. Beyond $*$, this 
category contains all objects 
of ${\cal A}$ which do not belong to $S$, and its morphism spaces are given by:
\bea
Hom_{{\cal A}_q}(*,*)&=&\oplus_{a,b\in S}{Hom_{\cal A}(a,b)}~~,\\
Hom_{{\cal A}_q}(a,*)&=&\oplus_{b\in S}{Hom_{\cal A}(a,b)}~{\rm~for~}a{\rm~not~in~}S~~,\\
Hom_{{\cal A}_q}(*,b)&=&\oplus_{a\in S}{Hom_{\cal A}(a,b)}~{\rm~for~}b{\rm~not~in~}S~~,\\
Hom_{{\cal A}_q}(a,b)&=&Hom_{\cal A}(a,b)~{\rm~for~}a,b{\rm~not~in~}S~~,
\eea
and with the morphism compositions induced from ${\cal H}$.
It is easy to see that $Q'$ preserves these subspaces of ${\cal H}$. 
We define the 
BRST operators on the new boundary sectors 
to be given by the restriction of $Q'$, while the bilinear forms 
are given by the restriction of $(.,.)$. 
For example, the BRST operator $Q_{**}$ 
acting on $Hom_{{\cal A}_q}(*,*)$ is given by:
\be
\label{Q_q}
(Q'_{**}u)=\oplus_{a,b\in S}{\left[
Q_{ab}u_{ab}+\sum_{c\in S}{\left(q_{cb}u_{ac}-(-1)^{|u_{cb}|}
u_{cb}q_{ac}\right)}\right]}~~,
\ee
for $u=\oplus_{a,b\in S}{u_{ab}}\in Hom(*,*)$, with $u_{ab}\in Hom(a,b)$. 
Its cohomology 
$H^*_{Q'_{**}}(Hom_{{\cal A}_q}(*,*))$ is the space of on-shell states of 
open strings in 
the new diagonal boundary sector produced by condensation of the operators $q_{ab}$. 
It is easy to see that  ${\cal A}_q$ endowed with this structure satisfies 
Axioms (1) and (2) of Section 2. This gives the description of our string field 
theory after shifting the vacuum.
The new decomposition into boundary sectors shows that
the pseudocomplex $q$ can be identified with the 
new object $*$, the D-brane 
composite obtained by condensing the boundary operators $(q_{ab})_{a,b\in S}$. 

\subsubsection{Shifts by reducible $q$}

If the complex is not connected, then each of the connected components 
$\Lambda$ 
of the associated graph $\Gamma$
defines a sub-collection $q_\Lambda=(q_{cd})_{c,d\in S_\Lambda}$, where $S_\Lambda$ is the 
set of objects of ${\cal A}$ associated with its vertices. In this case, the sets 
$S_\Lambda$ form a partition (disjoint union decomposition) of the set $S$, and it is easy 
to see that each collection $q_\Lambda$ is itself a pseudocomplex, i.e. it satisfies 
the tadpole cancellation constraint (\ref{tadpole_expanded}). 
Hence the reducible complex $q$ can be viewed as the collection of irreducible 
complexes $(q_\Lambda)_{\Lambda\in {\cal P}}$, where ${\cal P}$ is the set of 
connected components of $\Gamma$.

In this case, 
an analysis similar to that above leads to the conclusion that 
$q$ gives a collection of irreducible D-branes described by the connected complexes 
$q_\Lambda$. Each composite D-brane arises through condensation of the boundary operators 
in $q_\Lambda$, and these condensation processes are independent since there is no nonzero 
boundary operator which condenses between D-branes belonging to 
distinct subcomplexes $q_{\Lambda_1}$ and $q_{\Lambda_2}$. The associated category ${\cal A}_q$ 
is obtained by collapsing each of the sets $S_\Lambda$ to a different object $*_\Lambda$ (figure 2). 

\hskip 0.8 in
\begin{center} 
\scalebox{0.8}{\begin{picture}(0,0)%
\includegraphics{contraction.pstex}%
\end{picture}%
\setlength{\unitlength}{4144sp}%
\begingroup\makeatletter\ifx\SetFigFont\undefined%
\gdef\SetFigFont#1#2#3#4#5{%
  \reset@font\fontsize{#1}{#2pt}%
  \fontfamily{#3}\fontseries{#4}\fontshape{#5}%
  \selectfont}%
\fi\endgroup%
\begin{picture}(6975,2996)(91,-2536)
\put(5761,-61){\makebox(0,0)[lb]{\smash{\SetFigFont{14}{16.8}{\familydefault}{\mddefault}{\updefault}
$Hom_{{\cal A}_q}(a,*_\Sigma)$
}}}
\put(5761,-1816){\makebox(0,0)[lb]{\smash{\SetFigFont{14}{16.8}{\familydefault}{\mddefault}{\updefault}
$Hom_{{\cal A}_q}(a,*_\Lambda)$
}}}
\put(4546,-2536){\makebox(0,0)[lb]{\smash{\SetFigFont{17}{20.4}{\familydefault}{\mddefault}{\updefault}
$*_\Lambda$
}}}
\put(4546,254){\makebox(0,0)[lb]{\smash{\SetFigFont{17}{20.4}{\familydefault}{\mddefault}{\updefault}
$*_\Sigma$
}}}
\put(2836,-1051){\makebox(0,0)[lb]{\smash{\SetFigFont{17}{20.4}{\familydefault}{\mddefault}{\updefault}
$a$
}}}
\put( 91,-106){\makebox(0,0)[lb]{\smash{\SetFigFont{17}{20.4}{\familydefault}{\mddefault}{\updefault}
$\Sigma$
}}}
\put( 91,-2221){\makebox(0,0)[lb]{\smash{\SetFigFont{17}{20.4}{\familydefault}{\mddefault}{\updefault}
$\Lambda$
}}}
\put(7066,-1096){\makebox(0,0)[lb]{\smash{\SetFigFont{17}{20.4}{\familydefault}{\mddefault}{\updefault}
$a$
}}}
\put(1981,-1456){\makebox(0,0)[lb]{\smash{\SetFigFont{14}{16.8}{\familydefault}{\mddefault}{\updefault}
$\oplus_{\scriptsize {\begin{array}{c}a\in S_\Lambda\\b\in S_\Sigma\end{array}}}{Hom_{{\cal A}}(a,b)}$
}}}
\put(4141,-1501){\makebox(0,0)[lb]{\smash{\SetFigFont{14}{16.8}{\familydefault}{\mddefault}{\updefault}
$Hom_{{\cal A}_q}(*_\Lambda,*_\Sigma)$
}}}
\put(2116, 29){\makebox(0,0)[lb]{\smash{\SetFigFont{14}{16.8}{\familydefault}{\mddefault}{\updefault}
$\oplus_{b\in S_\Sigma}{Hom_{{\cal A}}(a,b)}$
}}}
\end{picture}
}
\end{center}
\begin{center} 
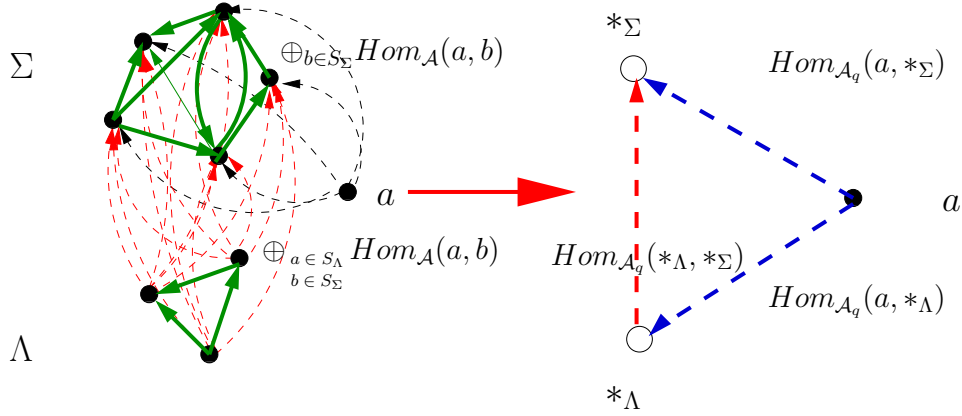
Figure  2. {\footnotesize Formation of two irreducible D-brane composites. After performing the 
vacuum shift, each of the two connected subgraphs on the left (the  arrows of which 
represent a morphism of the associated complex $q$) is replaced by a single 
object (the two unfilled circles on the right). 
We also draw one of the objects which does not belong 
to the associated complex, and show how the various morphism 
spaces involving D-brane 
composites are built from the original morphisms (the dashed lines). We do not 
indicate the morphisms making up $Hom(a,*_\Sigma)$, in order to prevent 
excessive clutter of the figure.}
\end{center}

The formal definition of ${\cal A}_q$ is as follows. The objects of 
${\cal A}_q$ are those objects of ${\cal A}$ which do not belong to $S$, 
together with the new objects $*_\Lambda$ ($\Lambda\in {\cal P}$).
Its morphism spaces are given by:  
\bea
\label{H_gen_bcc}
Hom_{{\cal A}_q}(*_\Lambda,*_\Sigma)&=&
\oplus_{a\in S_\Lambda, b\in S_\Sigma}{Hom_{\cal A}(a,b)}~~,\nn\\
Hom_{{\cal A}_q}(*_\Lambda,b)~~~&=&
\oplus_{a\in S_\Lambda}{Hom_{\cal A}(a,b)}~~{\rm for~}b{\not \in} S~~,\\
Hom_{{\cal A}_q}(a,*_\Sigma)~~&=&
\oplus_{b\in S_\Sigma}{Hom_{\cal A}(a,b)}~~{\rm~for~}a{\not \in} S~~,\nn\\
Hom_{{\cal A}_q}(a,b)~~~~&=& Hom_{\cal A}(a,b)~~{\rm~for~}a,b{\not \in} S~~,\nn
\eea
with the grading induced from that on $Hom(a,b)$. Morphism compositions 
are given by restriction of the associative composition in ${\cal H}$.
The BRST operator (\ref{Q'_expanded}) induces the BRST charges in these new boundary 
sectors:
\bea
\label{Q_gen_bcc}
Q'_{*_\Lambda, *_\Sigma}u&=&\oplus_{a\in S_\Lambda, b\in S_\Sigma}{\left[
Q_{ab}u_{ab}+\sum_{c\in S_\Sigma}{q_{cb}u_{ac}}-
\sum_{c\in S_\Lambda}{(-1)^{|u_{cb}|}u_{cb}q_{ac}}\right]}~~,\nn\\
Q'_{a, *_\Sigma}u&=&\oplus_{b\in S_\Sigma}{\left[
Q_{ab}u_{ab}+\sum_{c\in S_\Sigma}{q_{cb}u_{ac}}\right]}~~(a{\not \in }S),\\
Q'_{*_\Lambda, b}u&=&\oplus_{a\in S_\Lambda}{\left[
Q_{ab}u_{ab}-
\sum_{c\in S_\Lambda}{(-1)^{|u_{cb}|}u_{cb}q_{ac}}\right]}~~(b{\not \in} S),\nn\\
Q'_{ab}~~&=&Q'_{ab}~~(a,b {\not \in} S)~~.\nn
\eea
We shall call ${\cal A}_q$ the {\em contraction} of the category ${\cal A}$ 
along $q$.
The cohomology 
of $Q'_{*_\Lambda, *_\Sigma}$ on the space $Hom_{{\cal A}_q}(*_\Lambda, *_\Sigma)$ gives 
the on-shell state space of strings stretching between the composites $*_\Lambda$ and 
$*_\Sigma$. As above, 
we have induced bilinear forms (from restriction of $(.,.)$) 
and it is easy to see that ${\cal A}_q$ 
satisfies Axioms (1) and (2) of Section 2. We conclude that {\em the shifted 
string field theory is described by the contracted category ${\cal A}_q$}
(Figure 3).

\hskip 0.8 in
\begin{center} 
\scalebox{0.6}{\begin{picture}(0,0)%
\includegraphics{moduli.pstex}%
\end{picture}%
\setlength{\unitlength}{4144sp}%
\begingroup\makeatletter\ifx\SetFigFont\undefined%
\gdef\SetFigFont#1#2#3#4#5{%
  \reset@font\fontsize{#1}{#2pt}%
  \fontfamily{#3}\fontseries{#4}\fontshape{#5}%
  \selectfont}%
\fi\endgroup%
\begin{picture}(4389,2061)(2194,-2626)
\put(4096,-1096){\makebox(0,0)[lb]{\smash{\SetFigFont{17}{20.4}{\rmdefault}{\bfdefault}{\updefault}
${\cal A}$
}}}
\put(5626,-781){\makebox(0,0)[lb]{\smash{\SetFigFont{17}{20.4}{\rmdefault}{\bfdefault}{\updefault}
${\cal A}_q$
}}}
\put(3421,-2626){\makebox(0,0)[lb]{\smash{\SetFigFont{17}{20.4}{\rmdefault}{\bfdefault}{\updefault}
${\cal M}$
}}}
\end{picture}
}
\end{center}
\begin{center} 
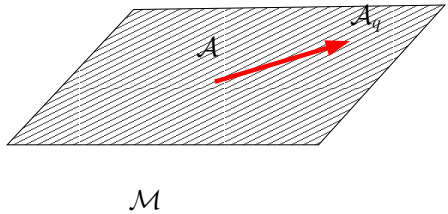
Figure  3. {\footnotesize Shifting the vacuum changes the description of 
our string field theory. The fact that the category structure changes 
signals the formation of D-brane composites.}
\end{center}

For reader's convenience, I list below the explicit 
morphism composition rules and bilinear forms 
on the category ${\cal A}_q$. 

1. {\bf Morphism compositions}
For $u\in Hom_{{\cal A}_q}(*_\Sigma, *_\Theta)$ and $v\in 
Hom_{{\cal A}_q}(*_\Lambda, *_\Sigma)$:
\be
uv=\oplus_{a\in S_\Lambda, b\in S_\Theta}{
\left[\sum_{c\in S_\Sigma}{u_{cb}v_{ac}}\right]}\in
Hom_{{\cal A}_q}(*_\Lambda, *_\Theta)~~.
\ee

For $u\in Hom_{{\cal A}_q}(*_\Sigma,c)$ and $v\in 
Hom_{{\cal A}_q}(*_\Lambda, *_\Sigma)$ ($c\not \in S$): 
\be
uv=\sum_{a\in S_\Lambda,b\in S_\Sigma}{u_bv_{ab}}\in Hom_{{\cal A}_q}(*_\Lambda,c)~~,
\ee
where $u=\sum_{b\in S_\Sigma}{u_b}$ with $u_b\in Hom_{\cal A}(b,c)$.

For $u\in Hom_{{\cal A}_q}(*_\Lambda, *_\Sigma)$, 
$v\in Hom_{{\cal A}_q}(a,*_\Lambda)$ ($a\not \in S$):
\be
uv=\oplus_{c\in S_\Sigma}{\left[\sum_{b\in S_\Lambda}{u_{bc}v_b}\right]}\in 
Hom_{{\cal A}_q}(a,*_\Sigma)~~,
\ee
where $v=\oplus_{b\in S_\Lambda}{v_b}$ with $v_b\in Hom_{\cal A}(a,b)$.

For $u\in Hom_{{\cal A}_q}(b,c)=Hom_{\cal A}(b,c)$ 
and $v\in Hom_{{\cal A}_q}(a,b)=Hom_{\cal A}(a,b)$, with $a,b \not \in S$b, 
the composition $uv$ is given by the morphism composition of ${\cal A}$. 

2. {\bf Bilinear forms}
For $u\in Hom_{{\cal A}_q}(*_\Sigma, *_\Lambda)$ and $v\in 
Hom_{{\cal A}_q}(*_\Lambda, *_\Sigma)$:
\be
(u,v)=\sum_{a\in S_\Sigma, b\in S_\Lambda}{(u_{ab}, v_{ba})}~~.
\ee

For $u\in Hom_{{\cal A}_q}(*_\Lambda,b)$ and $v\in 
Hom_{{\cal A}_q}(b, *_\Lambda)$ ($b\not \in S$): 
\be
uv=\sum_{a\in S_\Lambda}{(u_a, v_a)}~~,
\ee
where $u=\sum_{a\in S_\Lambda}{u_a}$ with $u_a\in Hom_{\cal A}(a,b)$
and   $v=\oplus_{a\in S_\Lambda}{u_a}$ with $u_a\in Hom_{\cal A}(b,a)$.

For $u\in Hom_{{\cal A}_q}(a, *_\Lambda)$, 
$v\in Hom_{{\cal A}_q}(*_\Lambda,a)$ ($a\not \in S$):
\be
uv=\sum_{b\in S_\Lambda}{(u_b,v_b)}~~.
\ee
where $u=\oplus_{b\in S_\Lambda}{u_b}$ with $u_b\in Hom_{\cal A}(a,b)$
and $v=\sum_{b\in S_\Lambda}{v_b}$ with $v_b\in Hom_{\cal A}(b,a)$.

For $u\in Hom_{{\cal A}_q}(a,b)=Hom_{\cal A}(a,b)$ 
and $v\in Hom_{{\cal A}_q}(b,a)=Hom_{\cal A}(b,a)$ with $a,b\not \in S$, 
$(u,v)$ is given by the bilinear form of ${\cal A}$. 

All of the sums involved in these expressions are well-defined if one restricts 
to pseudocomplexes whose underlying set of objects $S$ is finite. 

\

\subsection{The quasiunitarity constraint}

We are now ready to formulate precisely the unitarity constraint discussed in the 
introduction. 

{\bf Definition} We say that a string field theory based on a dG category is 
{\em quasiunitary} (or {\em closed under formation of D-brane composites}) if 
for any pseudocomplex $q$, the contracted category ${\cal A}_q$ is dG-equivalent 
with a full subcategory\footnote{Remember that a full subcategory ${\cal B}$ 
of a category ${\cal A}$ is a subcategory such that $Hom_{\cal B}(a,b)=Hom_{\cal A}(a,b)$
for any two objects $a,b\in {\cal B}$.}
of the category ${\cal A}$ via a functor which preserves the bilinear forms. 

Two dG categories are dG-equivalent if they are related by an equivalence of categories
which preserves the differentials on morphisms. A formal definition can be found in 
Appendix A.

This definition formalizes 
the intuition that performing a shift of the vacuum should not 
produce new boundary sectors. A `unitary' theory should be such that formation of 
D-brane composites 
always `reduces' the set of objects (D-brane) originally available. This encodes the basic 
constraint that the original theory is `complete', i.e. the state space is large enough 
in order to describe the dynamics. 

It is clear that a `generic' string field theory in the sense of Section 2 does not 
satisfy this constraint. For examples, topological B-type string theory on a Calabi-Yau 
manifold, in the presence of D-branes described by holomorphic vector bundles
satisfies Axioms (1) and (2) of Section 2 
but it is not unitary in this sense, since a shift of the 
vacuum leads to D-brane composites described by complexes of holomorphic vector bundles 
(and, in fact, to much more general objects, as we shall show in a companion paper). 
Similar remarks 
apply to physical string theories, such as superstring field theory on a Calabi-Yau 
manifold in the presence of D-branes. A theory which is not unitary at least in this 
weak sense is physically incomplete if viewed as a space-time 
description of D-brane dynamics, though it is of course well-defined as 
an abstract object. We propose that any open string field theory with D-branes 
should be formulated in a quasi-unitary manner.

\section{The quasi-unitary cover of an open string theory}

Given an open string field theory based on the category ${\cal A}$, it 
seems intuitively plausible 
that one could construct a quasi-unitary completion of that theory
by enlarging the class of objects through inclusion of all possible D-brane composites
($=$pseudocomplexes). 
It turns out that it is slightly subtle to give a precise formulation of 
this intuition, due to the following observation
({\em the problem of infinite recursion}).

The reason for expecting a quasiunitary completion is that one can consider
all possible D-brane composites and enlarge the original category ${\cal A}$ 
by adding them as new objects. It is in fact possible to formulate a category 
$p({\cal A})$ of all 
pseudocomplexes over ${\cal A}$ by simply using relations (\ref{H_gen_bcc}) and 
(\ref{Q_gen_bcc}) to define the morphism spaces between pseudocomplexes and the 
associated BRST operators, and composing morphisms through the obvious analogue 
of the composition in ${\cal A}_q$
\footnote{Since it certainly possible to perform {\em distinct} shifts 
of the vacuum leading to D-brane composites based on non-disjoint collections of 
objects of ${\cal A}$ (i.e. the associated pseudocomplexes share some of their 
objects), this involves a slight generalization.}. This category contains ${\cal A}$
as a full subcategory upon identifying objects of ${\cal A}$ with the associated 
pseudocomplex based on one object carrying the zero morphism. However, it 
turns out that the category 
$p({\cal A})$ is {\em not} closed under formation of D-brane composites. It follows 
that one has to repeat the process by building the category 
$p^2({\cal A})=p(p({\cal A}))$ of 
pseudocomplexes over $p({\cal A})$ (which corresponds to including composites 
arising by condensing boundary operators between various D-brane composites), 
then the categories $p^3({\cal A})$, $p^4({\cal A})$ and so forth. While this approach 
is certainly allowed, a quasiunitary completion would only be obtained as a limit
of $p^n({\cal A})$ when $n$ tends to infinity, and its explicit description seems 
difficult to extract. Moreover, one would have to define in what sense one 
takes such a limit (presumably an inductive limit).

Below, I show that this problem can be avoided by the deceivingly simple device 
of allowing for complexes based on {\em sequences} of objects of ${\cal A}$. 
This amounts to passing from pseudocomplexes to so-called {\em generalized 
complexes}, which are essentially pseudocomplexes whose underlying objects are 
allowed to be identical (since they are based on a sequence of objects rather 
than a set of objects, and a sequence may have repetitions). 
This apparently trivial extension suffices to `sum up' all of the 
recursive extensions $p^n({\cal A})$, as one can prove directly that the resulting 
category of generalized complexes is quasiunitary. Note, however, that a
generalized complex does not have a very direct interpretation as a D-brane composite, 
unless one agrees that it makes sense to condense multiple copies of a 
D-brane with itself. 

Before proceeding with a more detailed explanation of this construction, 
let me comment on the relation between our generalized complexes and 
apparently similar complexes of D-branes considered in the work of 
\cite{Oz_triples, Douglas_Kontsevich} 
(see also \cite{Oz_superconn} for related issues).  
In the approach of those papers, one looks at type IIB 
superstring theory on Calabi-Yau manifolds and obtains complexes 
based on sequences of holomorphic vector bundles. This allows for repetitions of 
the same bundle in a sequence, but such sequences do {\em not} correspond to a 
generalized complex of D-branes in the sense of our paper. 
The reason in that, in the approach of 
\cite{Oz_triples, Douglas_Kontsevich}, two consecutive copies of the same bundle 
in a sequence must correspond to a brane -antibrane pair, so that the associated 
D-branes are physically distinct even though the underlying holomorphic vector bundles
are identical. 
In a generalized complex as considered in this paper, one has multiple 
copies of the same {\em physical} D-brane, identified with an object of ${\cal A}$. 
As will be shown somewhere else, applying our construction to the open topological 
B-model gives (among much more general objects) sequences of holomorphic vector bundles (with repetitions allowed) with morphisms 
given by bundle-valued differential forms of arbitrary degree. 
This suffices to recover the complexes of 
degree zero morphisms considered in 
\cite{Oz_triples, Douglas_Kontsevich} and much more, but via a rather different 
procedure, which consists of considering the so-called shift completion of the 
category of holomorphic vector bundles. Moreover, note that our formalism 
applies perfectly well to associative bosonic string field theory, for which  
no simple analogue of the arguments of \cite{Oz_triples, Douglas_Kontsevich}
seems to exist\footnote{The reader 
has been warned that we are talking about something quite different from 
the brane-antibrane sequences of \cite{Oz_triples, Douglas_Kontsevich}, in spite 
of what a shallow analogy might suggest. To the extent that open superstring 
field theory satisfies our axioms (which is not entirely clear), the original 
category ${\cal A}$ would contain both branes and antibranes, which in our
treatment are distinct objects. We are interested in
sequences with repetitions of the same D-brane, and not in sequences of 
brane/antibrane pairs.}.

\subsection{Motivation}

The proposal below is based on our discussion of formation of D-brane composites. 
As mentioned above, a unitary description requires 
passage from pseudocomplexes to a more general construction (`generalized complexes') 
which involve  {\em sequences} of objects of ${\cal A}$. 
Before presenting the construction itself, let me outline the motivation for 
introducing such objects. Consider minimally 
extending the category ${\cal A}$ to a dG category ${\cal B}$ which satisfies the 
quasiunitarity constraint. It is clear that a minimal extension 
must amount at least to addition of all 
pseudocomplexes ($=$ D-brane composites). Hence the enlarged category ${\cal B}$ 
contains at least all pseudocomplexes based on 
the objects of ${\cal A}$. Note that an object $a$ of ${\cal A}$ can be identified 
with the one-morphism pseudocomplex $(0_a)$, where $0_a$ is the zero morphism in 
$Hom_{\cal A}(a,a)$. 

Now consider an object $a$ of ${\cal A}$ and a pseudocomplex $q=(q_{aa})$
over ${\cal A}$ with a single nonzero morphism $q_{aa}\in Hom^1(a,a)$ 
based on $a$. 
The tadpole cancellation constraint for such a complex reads: 
\be
\label{tadpole_q}
Q_{aa}q_{aa}+q_{aa}q_{aa}=0~~,
\ee
which is the Maurer-Cartan equation for vacuum deformations in the boundary sector 
$Hom_{\cal A}(a,a)$ 
(it follows that such  a complex describes deformations of the D-brane 
$a$ of ${\cal A}$). We shall make the assumption that the associated morphism 
spaces and BRST differentials in ${\cal B}$ are given by the generalization of 
(\ref{H_gen_bcc}) and (\ref{Q_gen_bcc}) to the case of pseudocomplexes 
based on non-disjoint sets of objects (we identify $a$ with the pseudocomplex $(0_a)$)
\footnote{The reason why this is only an assumption is that 
our description of D-brane condensation does not immediately tell us how to describe 
morphisms between pseudocomplexes whose underlying sets of objects are not 
disjoint. Assuming this form of morphisms and BRST operators amounts to 
the statement that any quasiunitary extension ${\cal B}$ of ${\cal A}$ contains 
the category of pseudocomplexes $p({\cal A})$ alluded to above.
This can be justified more formally but I prefer to avoid further technicalities.}:
\be
\label{hom_b}
Hom_{\cal B}(a,a)=Hom_{\cal B}(a,q)=Hom_{\cal B}(q,a)=Hom_{\cal B}(q,q)=Hom_{\cal A}(a,a)
\ee
and:
\bea
\label{q_b}
Q^{\cal B}_{aa}u&=&Q_{aa}u~~{\rm~for~}u\in Hom_{\cal B}(a,a)~~\nn\\
Q^{\cal B}_{aq}u&=&Q_{aa}u+q_{aa}u~~{\rm~for~}u\in Hom_{\cal B}(a,q)~~\nn\\
Q^{\cal B}_{qa}u&=&Q_{qa}u-(-1)^{|u|}uq_{aa}~~{\rm~for~}u\in Hom_{\cal B}(q,a)~~\\
Q^{\cal B}_{qq}u&=&Q_{aa}u+q_{aa}u-(-1)^{|u|}uq_{aa}~~{\rm~for~}u\in Hom_{\cal B}(q,q)~~.
\nn
\eea
This assumption is natural in view our previous discussion of D-brane condensation. 

Both the complex $q$ and the D-brane $a\equiv(0_a)$ are objects of the 
enlarged category ${\cal B}$, and one can consider condensation 
(in the extended string theory defined by ${\cal B}$) of 
boundary operators $f\in Hom_{\cal B}(a,q)$. 
Suppose that we pick such a morphism $f_{aq}\in Hom^1_{\cal B}(a,q)$ 
which defines a pseudocomplex $f=(f_{aq})$ in ${\cal B}$ 
over the objects $a$, $q$ of ${\cal B}$, i.e. such that the tadpole cancellation 
constraint in ${\cal B}$ is satisfied (see figure 4):
\be
\label{tadpole_f}
Q_{aq}^{\cal B}f_{aq}=0\Leftrightarrow Q_{aa}f_{aq}+q_{aa}f_{aq}=0~~.
\ee
Condensation of $f$ then produces a contracted category ${\cal B}_q$ which must be 
equivalent with a subcategory of ${\cal B}$ since ${\cal B}$ 
was assumed to be quasiunitary. In particular, ${\cal B}$ must contain an object 
$*_f$ such that $Hom_{\cal B}(*_f,*_f)=Hom_{{\cal B}_q}(*_f,*_f)$, i.e.:
{\footnotesize \bea
Hom_{\cal B}(*_f,*_f)=Hom_{\cal B}(a,a)\oplus Hom_{\cal B}(q,q)\oplus Hom_{\cal B}(a,q)
\oplus Hom_{\cal B}(q,a)=Hom_{\cal A}(a,a)^{\oplus 4}~~,\nn
\eea}\noindent where we used our assumptions (\ref{hom_b}). 
The BRST charge on this space is:
{\footnotesize\bea
& &Q^{\cal B}_{*_f,*_f}u=Q^{{\cal B}_q}_{*_f,*_f}u=
[Q^{\cal B}_{aa}u_{aa}-(-1)^{|u_{qa}|}u_{qa}f_{aq}]\oplus \nn\\
& \oplus & [Q^{\cal B}_{qq}u_{qq}+f_{aq}u_{qa}]
\oplus 
[Q^{\cal B}_{aq}u_{aq} -(-1)^{|u_{qq}|}u_{qq}f_{aq}]
\oplus 
[Q^{\cal B}_{qa}u_{qa}]
\eea}\noindent for $u=u_{aa}\oplus u_{qq}\oplus u_{aq}\oplus u_{qq}$ with 
$u_{ef}\in Hom_{\cal B}(e,f)=Hom_{\cal A}(a,a)$ for all $e,f\in \{a,q\}$. 
In view of our assumptions (\ref{q_b}), this equals:
{\footnotesize
\bea
\label{Qstarf}
Q^{\cal B}_{*_f,*_f}u &=&[Q_{aa}u_{aa}-(-1)^{|u_{qa}|}u_{qa}f_{aq}]\oplus 
[Q_{aa}u_{qq}+q_{aa}u_{qq}-(-1)^{|u_{qq}|}u_{qq}q_{aa}+f_{aq}u_{qa}]
\oplus \nn\\
& \oplus &[Q_{aa}u_{aq} +q_{aa}u_{aq}-(-1)^{|u_{qq}|}u_{qq}f_{aq}]
\oplus 
[Q_{qa}u_{qa}-(-1)^{|u_{qa}|}u_{qa}q_{aa}]~~.
\eea}

\hskip 0.1 in
\begin{center} 
\scalebox{0.5}{\begin{picture}(0,0)%
\includegraphics{f.pstex}%
\end{picture}%
\setlength{\unitlength}{4144sp}%
\begingroup\makeatletter\ifx\SetFigFont\undefined%
\gdef\SetFigFont#1#2#3#4#5{%
  \reset@font\fontsize{#1}{#2pt}%
  \fontfamily{#3}\fontseries{#4}\fontshape{#5}%
  \selectfont}%
\fi\endgroup%
\begin{picture}(8517,3960)(1,-3121)
\put(1891,-2806){\makebox(0,0)[lb]{\smash{\SetFigFont{17}{20.4}{\familydefault}{\mddefault}{\updefault}
$*_f$
}}}
\put(3376,-2806){\makebox(0,0)[lb]{\smash{\SetFigFont{17}{20.4}{\familydefault}{\mddefault}{\updefault}
$\equiv$
}}}
\put(4591,-2806){\makebox(0,0)[lb]{\smash{\SetFigFont{17}{20.4}{\familydefault}{\mddefault}{\updefault}
$c=$
}}}
\put(6301,-151){\makebox(0,0)[lb]{\smash{\SetFigFont{17}{20.4}{\familydefault}{\mddefault}{\updefault}
;
}}}
\put(7516,-196){\makebox(0,0)[lb]{\smash{\SetFigFont{17}{20.4}{\familydefault}{\mddefault}{\updefault}
$=$
}}}
\put(1621,569){\makebox(0,0)[lb]{\smash{\SetFigFont{17}{20.4}{\familydefault}{\mddefault}{\updefault}
(in ${\cal B}$)
}}}
\put(8191,-2716){\makebox(0,0)[lb]{\smash{\SetFigFont{17}{20.4}{\familydefault}{\mddefault}{\updefault}
(in ${\cal B}$)
}}}
\put(  1,-2806){\makebox(0,0)[lb]{\smash{\SetFigFont{17}{20.4}{\familydefault}{\mddefault}{\updefault}
(in ${\cal B}_f$)
}}}
\put(5851,-2536){\makebox(0,0)[lb]{\smash{\SetFigFont{17}{20.4}{\familydefault}{\mddefault}{\updefault}
$q_{12}=f_{aq}$
}}}
\put(5356,-3076){\makebox(0,0)[lb]{\smash{\SetFigFont{17}{20.4}{\familydefault}{\mddefault}{\updefault}
$a_1=a$
}}}
\put(7471,-3121){\makebox(0,0)[lb]{\smash{\SetFigFont{17}{20.4}{\familydefault}{\mddefault}{\updefault}
$a_2=a$
}}}
\put(7156,-1861){\makebox(0,0)[lb]{\smash{\SetFigFont{17}{20.4}{\familydefault}{\mddefault}{\updefault}
$q_{22}=q_{aa}$
}}}
\put(8281,-556){\makebox(0,0)[lb]{\smash{\SetFigFont{17}{20.4}{\familydefault}{\mddefault}{\updefault}
$a$
}}}
\put(8146,659){\makebox(0,0)[lb]{\smash{\SetFigFont{17}{20.4}{\familydefault}{\mddefault}{\updefault}
$q_{aa}$
}}}
\put(3016,-646){\makebox(0,0)[lb]{\smash{\SetFigFont{17}{20.4}{\familydefault}{\mddefault}{\updefault}
$q$
}}}
\put(946,-646){\makebox(0,0)[lb]{\smash{\SetFigFont{17}{20.4}{\familydefault}{\mddefault}{\updefault}
$a$
}}}
\put(1801, 29){\makebox(0,0)[lb]{\smash{\SetFigFont{17}{20.4}{\familydefault}{\mddefault}{\updefault}
$f_{aq}$
}}}
\put(7111,-601){\makebox(0,0)[lb]{\smash{\SetFigFont{17}{20.4}{\familydefault}{\mddefault}{\updefault}
$q$
}}}
\end{picture}
}
\end{center}
\begin{center} 
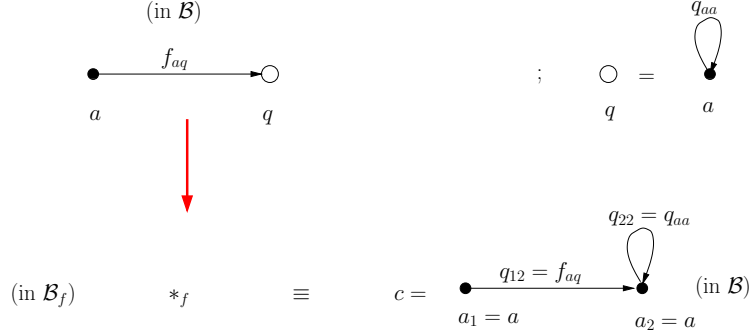
Figure  4. {\footnotesize The pseudocomplex $f$ over ${\cal B}$ and the 
associated generalized complex $c$ over ${\cal A}$. 
In the upper left corner, we show $f$ as 
a pseudocomplex over ${\cal B}$. The associated vacuum shift (condensation of the 
boundary operator $f_{aq}$ between the objects $a$ and $q$ of ${\cal B}$) 
produces an object $*_f$ in the collapsed category 
${\cal B}_f$, which is identified with the 
generalized complex $c$ over ${\cal A}$. 
The latter can be viewed as an element of ${\cal B}$, 
if ${\cal B}$ is pseudounitary extension of ${\cal A}$. This type of construction 
can be used to argue that a quasiunitary extension 
${\cal B}$ of ${\cal A}$ contains all generalized 
complexes over ${\cal A}$. The introduction of generalized complexes amounts 
intuitively to allowing for the formation of composites 
between a D-brane and its deformations.}
\end{center}

It is clear that the object $*_f$ cannot be identified with a pseudocomplex $s$
over ${\cal A}$, since the morphism space $Hom(s,s)$ for a pseudocomplex cannot 
contain a direct power of the form $Hom(a,a)^{\oplus 4}$ (this is due to the fact that 
pseudocomplexes are based on {\em sets} of objects of ${\cal A}$, and 
thus cannot contain repetitions of the same object). Hence a minimal quasiunitary 
extension ${\cal B}$ must consist of more than pseudocomplexes over 
${\cal A}$. In fact, the object $*_f$ can be represented by a so-called generalized 
complex, namely by a {\em sequence} of objects and degree one morphisms of ${\cal A}$
satisfying a generalization of the tadpole cancellation constraint (a formal 
definition of generalized complexes is given below). For this, we define:
\bea
a_1=a_2=a~~&,&~~q_{11}=q_{21}=0~~\\
q_{12}=f_{aq}~~&,&~~q_{22}=q_{aa}~~,
\eea
and view this data as a two-term sequence $(a_1,a_2)$ of objects of ${\cal A}$ 
together with the degree one morphisms $q_{ij}$($i,j=1,2$). It is easy to see 
that the tadpole cancellation constraints (\ref{tadpole_q}) and (\ref{tadpole_f}) 
are  equivalent with:
\be
Q_{a_ia_j}q_{ij}+\sum_{k}{q_{kj}q_{ik}}=0~~.
\ee
Hence $c:=[(a_i)_{i=1,2}, (q_{ij})_{i,j=1,2}]$ is a degree one generalized complex 
over ${\cal A}$ in the sense defined below (this is depicted in Figure 4). 
It is shown below that generalized complexes 
form a dG category $c({\cal A})$, in which the endomorphism space of $c$
is:
\be
Hom_{c({\cal A})}(c,c)=\oplus_{i,j=1}^2{Hom_{\cal A}(a_i,a_j)}=Hom(a,a)^{\oplus 4}~~,
\ee
with the BRST charge:
\be
Q_{cc}u=\oplus_{i,j=1}^2{
\left[Q_{a_i,a_j}u_{ij}+\sum_{k}{q_{kj}u_{ik}}-\sum_{k}{(-1)^{|u_{kj}|}u_{kj}q_{ik}}\right]}
\ee
for $u=\oplus_{i,j=1,2}{u_{ij}}$ with $u_{ij}\in Hom(a_i,a_j)=Hom(a,a)$. It is 
easy to see that this coincides with (\ref{Qstarf}). 

This simple example explains the need for introducing generalized complexes, and 
can be extended to show that in a certain sense the category 
of generalized complexes $c({\cal A})$ constructed below gives the 
{\em minimal} 
quasiunitary extension of the string field theory based on ${\cal A}$.

\subsection{Formal definition}

We now proceed to give a formal definition of generalized complexes and of the category $c({\cal A})$.

{\bf Definition} A (degree one) 
{\em generalized complex} over ${\cal A}$ is given by the following data:

(1) a {\em sequence} $(a_i)_{i\in I}$ of objects of ${\cal A}$ (note that the 
objects $a_i$ {\em need not be distinct}), where $I$ is some finite set of indices. 

(2) for each $i,j\in I$, a morphism $q_{ij}\in Hom(a_i,a_j)$. These morphisms 
are subject to the constraints:

\be
Qq_{ij}+\sum_{k\in I}{q_{kj}q_{ik}}=0~~.
\ee

\

We say that a generalized complex has degree $k$ if $q_{ij}\in Hom^k(a_i,a_j)$ for all 
$i,j\in I$.

{\bf Definition} Given a dG category ${\cal A}$, its {\em quasiunitary cover} 
is the dG category $c({\cal A})$ constructed as follows.

(1) The objects of $c({\cal A})$ are generalized complexes of degree one.

(2) If $q=[(a_i)_{i\in I},(q_{ij})_{i,j\in I}]$ and 
$q'=[(a'_i)_{i\in J},(q'_{ij})_{i,j\in J}]$ are two generalized complexes of degree one, 
then the space of morphisms $Hom_{c({\cal A})}(q,q')$ is given by:
\be
Hom(q,q')=\oplus_{i\in I,j\in J}{Hom(a_i,a'_j)}~~,
\ee
(this is inspired by relation (\ref{H_gen_bcc}) and our previous example).

(3) Given a third degree one generalized complex $q''=[(a''_i)_{i\in K}, 
(q''_{ij})_{i,j\in K}]$, 
the composition of morphisms $Hom(q',q'')\times Hom(q,q')\rightarrow Hom(q,q'')$ 
is given by:
\be
uv=\oplus_{i\in I, k\in K}{\left[\sum_{j\in J}{u_{jk}v_{ij}}\right]}~~,
\ee
where $v=\oplus_{i\in I, j\in J}{v_{ij}}\in Hom(q,q'), u=\oplus_{j\in J, k\in K}
{u_{jk}}\in Hom(q', q'')$, with $v_{ij}\in Hom(a_i,a'_j)$ and $u_{jk}\in Hom(a'_j,a''_k)$. 

(4) The BRST operator $Q_{qq'}:Hom(q,q')\rightarrow Hom(q,q')$ is defined via the 
analogue of (\ref{Q_gen_bcc}):
\be
Q_{qq'}u=\oplus_{i\in I, j\in J}{\left[
Q_{a_i a'_j}u_{ij}+\sum_{k\in J}{q'_{kj}u_{ik}}-
\sum_{l\in I}{(-1)^{|u_{lj}|}u_{lj}q_{il}}\right]}
\ee
for $u=\oplus_{i\in I, j\in J}{u_{ij}}\in Hom(q,q')$, with $u_{ij}\in Hom(a_i,a'_j)$.

(5)The grading on $Hom(q,q')$ is induced from the grading on the spaces 
$Hom(a,b)$, i.e.:
\be
Hom^k(q,q')=\oplus_{i\in I,j\in J}{Hom^k(a_i,a'_j)}~~.
\ee

\

It is clear that a pseudocomplex $q=(q_{ab})_{a,b\in S}$ can be viewed as 
a (degree one) 
generalized complex upon choosing an enumeration $S=\{a_i|i \in I\}$ of its underlying set 
of objects (note that there are many such generalized complexes, differing by a choice 
in the enumeration of $S$). However, not every generalized complex is of this type, since for 
a generalized complex $q=[(a_i)_{i\in I}, (q_{ij})_{i,j\in I}]$ one can have repetitions 
in the sequence of objects $(a_i)_{i\in I}$. For example, all objects $a_i$ could 
be identical, $a_i=a$ for all $i$. In this sense, a (degree one) 
generalized complex is a generalization 
of a pseudocomplex, and cannot be obtained in a simple fashion by performing a 
vacuum deformation of the original string field theory. This is, in particular, why 
passage to $c({\cal A})$ gives something essentially new. 

It is not hard to check that $c({\cal A})$ is a dG category. The only slightly nontrivial 
statement is that the operators $Q_{qq'}$ are nilpotent. This results from a computation which 
I shall leave as an exercise for the diligent reader. Moreover, we can endow 
$c({\cal A})$ with bilinear forms $(.,.):Hom(q',q)\times Hom(q,q')\rightarrow \C$,
defined as follows:
\be
(u,v)=\sum_{i\in I, j\in J}{(u_{ji},v_{ij})}
\ee
for $u=\oplus_{j\in J,i\in I}{u_{ji}}$, $v=\oplus_{i\in I, j\in J}{v_{ij}}$ with 
$u_{ji}\in Hom(a'_j,a_i)$ and 
$v_{ij}\in Hom(a_i,a'_j)$. It is easy to check that the category 
$c({\cal A})$ endowed with these forms  satisfies Axioms (1) and (2) of Section 2 
and hence defines a string field theory, which we shall call the {\em 
quasiunitary completion} of the string field theory based on ${\cal A}$
\footnote{It can be argued that 
this extension is minimal in the sense that any other extension
contains it. This requires a slightly technical analysis which 
amounts to showing that 
any quasiunitary extension ${\cal B}$ of ${\cal A}$ must include all 
generalized degree one complexes over ${\cal A}$. This is a generalization of
our the example we considered in Subsection 5.1}. 

\subsection{Quasiunitarity of $c({\cal A})$} 

We now proceed to show that $c({\cal A})$ gives a quasiunitary string theory in 
the sense of the previous subsection. The argument given below is pretty straightforward, 
but slightly involved, and can be skipped at a first reading of the paper. 

\

{\bf Proposition} The string field theory $c({\cal A})$ is quasi-unitary. 

\

{\bf Proof~}
Let $f$ be a  degree one pseudocomplex over $c({\cal A})$. We have to show
that the contracted category ${\cal B}=c({\cal A})_f$ is dG-equivalent with a 
subcategory of $c({\cal A})$ (see Appendix 1 for 
a precise definition of the notion of dG-equivalence). 
For simplicity, we assume that $f$ is connected (it is easy to see 
that it suffices to check 
the statement for connected complexes\footnote{This follows from the observation that 
condensing two disjoint connected pseudocomplexes $q$, $q'$ simultaneously (which is the same 
as condensing the disconnected pseudocomplex given by their union) is equivalent with 
first condensing $q$ and then condensing $q'$ in the category contracted along $q$.One can alternately 
extend the computations below to the case of disconnected pseudocomplexes, at the 
price of doubling the length of some formulas. }. 
We start by describing the 
contracted category ${\cal B}$.

\

\noindent{\em Description of ${\cal B}$}

\

Since $f$ is a degree one pseudocomplex over 
$c({\cal A})$, it is in particular 
a degree one generalized complex over $c({\cal A})$, provided that we chose an 
enumeration of its set of objects, and 
we assume that such an enumeration has been chosen. Then $f=[(q^{(\alpha)})_{
\alpha \in A}, (f^{(\alpha\beta)})_{\alpha\beta\in A}]$ for some set $A$, with 
$q^{(\alpha)}$ some degree one generalized complexes 
over ${\cal A}$ (i.e. objects of 
$c({\cal A})$) and
$f^{(\alpha\beta)}\in Hom^1_{c({\cal A})}(q^{(\alpha)},q^{(\beta)})$, satisfying the 
tadpole cancellation constraint for $c({\cal A})$:
\be
\label{moo}
Q^{(\alpha\beta)}f^{(\alpha\beta)}+\sum_{\gamma\in A}
{f^{(\gamma\beta)}f^{(\alpha\gamma)}}=0~~.
\ee
Here $Q^{(\alpha\beta)}$ is the BRST operator on 
$Hom_{c({\cal A})}(q^{(\alpha)}, q^{(\beta)})$.

Now, each $q^{(\alpha)}=[(a^{(\alpha)}_i)_{i\in I_\alpha}, (q^{(\alpha)}_{ij})_{i,j\in I_\alpha}]$ is 
a degree one generalized complex over ${\cal A}$, with 
$(a^{(\alpha)}_i)_{i\in I_\alpha}$ a family of objects of ${\cal A}$ and 
$q^{(\alpha)}_{ij}\in Hom^1_{\cal A}(a^{(\alpha)}_i, a^{(\alpha)}_j)$. 
Moreover, each $f$ is a family of morphisms in the category ${\cal A}$:
\be
f^{(\alpha\beta)}=
\oplus_{i\in I_\alpha, j\in I_\beta}
{f^{(\alpha\beta)}_{ij}}~~{\rm~with~}f^{(\alpha\beta)}_{ij}\in Hom^1_{\cal A}(a^{(\alpha)}_i,
a^{(\beta)}_j)~~,
\ee
and we have:
\be
Q^{(\alpha\beta)}f^{(\alpha\beta)}=\oplus_{i\in I_\alpha, j\in I_\beta}
{\left[Q_{a^{(\alpha)}_ia^{(\beta)}_j}f^{(\alpha\beta)}_{ij}+
\sum_{k\in I_\beta}{q^{(\beta)}_{kj}f^{(\alpha\beta)}_{ik}}+
\sum_{k\in I_\alpha}{f^{(\alpha\beta)}_{kj}q^{(\alpha)}_{ik}}\right]}~~
\ee
(where we used the fact that all components of $f$ and $q$ have degree one)
and:
\be
f^{(\beta\gamma)}f^{(\alpha\beta)}=\oplus_{i\in I_\alpha,j\in I_\beta}{
\left[\sum_{k\in I_\gamma}{f^{(\beta\gamma)}_{kj}f^{(\alpha\beta)}_{ik}}\right]}~~. 
\ee
Hence the tadpole cancellation condition (\ref{moo}) reduces to:
\be
\label{tad_f}
Q_{a^{(\alpha)}_i,a^{(\beta)}_j}f^{(\alpha\beta)}_{ij}+
\sum_{k\in I_\beta}{q^{(\beta)}_{kj}f^{(\alpha\beta)}_{ik}}+
\sum_{k\in I_\alpha}{f^{(\alpha\beta)}_{kj}q^{(\alpha)}_{ik}}+
\sum_{\gamma \in A, k\in I_\gamma}{f^{(\gamma\beta)}_{kj}f^{(\alpha\gamma)}_{ik}}=0~~.
\ee
On the other hand, we have the tadpole cancellation constraints obeyed by each of the 
complexes $q^{(\alpha)}$:
\be
\label{tad_q}
Q_{a^{(\alpha)}_i, a^{(\alpha)}_j}q^{(\alpha)}_{ij}+
\sum_{k\in I_\alpha}{q^{(\alpha)}_{kj}q^{(\alpha)}_{ik}}=0~~. 
\ee

Consider now the contracted category 
${\cal B}=c({\cal A})_f$. Its objects are those objects 
of $c({\cal A})$ which differ from all $q^{(\alpha)}$, plus a new objects $*_f$. 
Its morphism spaces are defined as above, and it sufficed to consider:
\be
\label{ff}
Hom_{\cal B}(f,f)=\oplus_{\alpha,\beta\in A}Hom_{c({\cal A})}(q^{(\alpha)},q^{(\beta)})
=\oplus_{\alpha,\beta\in A}\oplus_{i\in I_\alpha, j\in I_\beta}{
Hom_{\cal A}(a^{(\alpha)}_i, a^{(\beta)}_j)}~~,
\ee
and 
\be
\label{qf}
Hom_{\cal B}(q,f)=\oplus_{\alpha\in A}{Hom_{c({\cal A})}(q, q^{(\alpha)})}=
\oplus_{\alpha\in A}\oplus_{i\in J, j\in I_\alpha}{
Hom_{\cal A}(b_i, a^{(\alpha)}_j)}~~,
\ee
where $q=[(b_i)_{i\in J}, (q_{ij})_{i,j\in J})]$ is an object of $c({\cal A})$
which differs from all $q^{(\alpha)}$. 

We shall need the BRST differentials on these spaces. On $Hom_{\cal B}(f,f)$ we have 
the differential $Q_{ff}$ given by:
\be
Q_{ff}{u}=\oplus_{\alpha\beta\in A}{\left[Q^{(\alpha\beta)}u^{(\alpha\beta)}+
\sum_{\gamma\in A}{f^{(\gamma\beta)}u^{(\alpha\gamma)}}-
\sum_{\gamma\in A}{(-1)^{|u^{(\gamma\beta)}|}
u^{(\gamma\beta)}f^{(\alpha\gamma)}}\right]}~~,
\ee
where $u=\oplus_{\alpha\beta}{u^{(\alpha\beta)}}\in Hom_{\cal B}(f,f)$, 
with $u^{(\alpha\beta)}=\oplus_{i\in I_\alpha, j\in I_\beta}{u^{(\alpha\beta)}_{ij}}\in 
Hom_{c({\cal A})}(q^{(\alpha)}, q^{(\beta)})$
and $u^{(\alpha\beta)}_{ij}\in Hom_{\cal A}(a^{(\alpha)}_i,a^{(\beta)}_j)$. This has the 
expansion:
{\tiny\bea
\label{Q_{ff}}
& &Q_{ff}u=\oplus_{\alpha\beta\in A}\oplus_{i\in I_\alpha,j\in I_\beta}~~\\
& &\left[Q_{a^{(\alpha)}_ia^{(\beta)}_j}u^{(\alpha\beta)}_{ij}+
\sum_{k\in I_\beta}{q^{(\beta)}_{kj}u^{(\alpha\beta)}_{ik}}-
\sum_{l\in I_\alpha}{(-1)^{|u^{(\alpha\beta)}_{lk}|}u^{(\alpha\beta)}_{lk}
q^{(\alpha)}_{il}}
+\sum_{\gamma\in A, k\in I_\gamma}{f^{(\gamma\beta)}_{kj}
u^{(\alpha\gamma)}_{ik}}-
\sum_{\gamma\in A, k\in I_\gamma}{(-1)^{|u^{(\gamma\beta)}_{kj}|}
u^{(\beta)}_{kj}f^{(\alpha\gamma)}_{ik}}
\right]~~.~~~~~~~~~~~~~~~~~~~~\nn
\eea}

On $Hom_{\cal B}(q, f)$, we have the operator:
\be
Q_{qf}u=\oplus_{\alpha\in A}{\left[Q_{qq^{(\alpha)}}u^{(\alpha)}
+\sum_{\beta\in A}{f^{(\beta\alpha)}u^{(\beta)}}\right]}~~,
\ee
for $u=\oplus_{\alpha\in A}{u^{(\alpha)}}$ with $u^{(\alpha)}\in 
Hom_{c({\cal A})}(q, q^{(\alpha)})$. Upon expanding $u^{(\alpha)}=
\oplus_{i\in J, j\in I_\alpha}{u^{(\alpha)}_{ij}}$, with $u^{(\alpha)}_{ij}\in 
Hom(b_i, a^{(\alpha)}_j)$, we have:
\be
Q_{qq^{(\alpha)}}u^{(\alpha)}=\oplus_{i\in J, j\in I_\alpha}{
\left[
Q_{b_i, a^{(\alpha)}_j}u^{(\alpha)}_{ij}+
\sum_{k\in I_\alpha}{q^{(\alpha)}_{kj}u^{(\alpha)}_{ik}}
-\sum_{k\in J}{(-1)^{|u^{(\alpha)}_{kj}|}u^{(\alpha)}_{kj}q_{ik}}
\right]}~~,
\ee
and
\be
f^{(\beta\alpha)}u^{(\beta)}=\oplus_{i\in J,j\in I_\alpha}{\left[
\sum_{k\in I_\beta}{f^{(\beta\alpha)}_{kj}u^{(\beta)}_{ik}}
\right]}~~.
\ee
Hence we can expand:
{\scriptsize \be
\label{Q_{qf}}
Q_{qf}u=
\oplus_{\alpha\in A}\oplus_{i\in J, j\in I_\alpha}{\left[
Q_{b_ia^{(\alpha)}_j}u^{(\alpha)}_{ij}+
\sum_{k\in I_\alpha}{q^{(\alpha)}_{kj}u^{(\alpha)}_{ik}}-
\sum_{k\in J}{(-1)^{|u^{(\alpha)}_{kj}|}u^{(\alpha)}_{kj}q_{ik}}+
\sum_{\beta \in A, k\in I_\beta}{f^{(\beta\alpha)}_{kj}u^{(\beta)}_{ik}}
\right]}~~.
\ee}

\

\noindent{\em The object of $c({\cal A})$ associated with $f$}

\

We now construct a degree one generalized complex ${\bf q}$ 
over ${\cal A}$ which can be identified 
with the generalized complex of generalized complexes given by $f$. We let 
$I=\sqcup_{\alpha\in A}{I_\alpha}$ be the disjoint union of the index sets $I_\alpha$. 
Remember that $I$ can be identified with the set of all pairs $(\alpha,i)$ with 
$\alpha\in A$ and $i\in I_\alpha$. We consider the family of objects of ${\cal A}$
indexed by $I$ and defined through $a(\alpha,i)=a^{(\alpha)}_i$ for all $\alpha\in A$ 
and $i\in I_\alpha$. Next, we consider the morphisms 
${\bf q}_{\alpha i, \beta j}\in Hom^1(a_{\alpha i}, a_{\beta j})=Hom^1(a^{(\alpha)}_i, 
a^{(\beta)}_j)$ given by:
\bea
\label{q_new}
{\bf q}_{\alpha i, \alpha j}&=&f^{(\alpha\alpha)}_{ij}+q^{(\alpha)}_{ij}~~\\
{\bf q}_{\alpha i\beta j}&=&f^{(\alpha\beta)}_{ij}~{\rm~for~}\alpha\neq \beta~~.
\eea
It is not hard to see that ${\bf q}$ satisfies the tadpole cancellation constraint 
of ${\cal A}$, namely:
\be
Q_{a^{(\alpha)}_i, a^{(\beta)}_j}{\bf q}_{\alpha i, \beta j}+
\sum_{\gamma\in A, k\in I_\gamma}{{\bf q}_{\gamma k\beta j}{\bf q}_{\alpha i \gamma k}}
=0~~.
\ee
This follows by using the tadpole cancellation constraints (\ref{tad_f}) for the 
generalized complex $f$ and (\ref{tad_q}) for the generalized complexes $q^{(\alpha)}$
and is left as an exercise for the reader. 

The next step is to compute the Hom spaces of ${\bf q}$ with itself and 
with with degree one 
generalized complexes $q=[(b_i)_{i\in J}, (q_{ij})_{i,j\in J}]$ of the type considered 
above. We have:
\be
Hom_{c({\cal A})}({\bf q}, {\bf q})=\oplus_{\alpha,\beta\in A}
\oplus_{i\in I_\alpha, j\in I_\beta}{Hom_{\cal A}(a^{(\alpha)}_i, a^{(\alpha)}_j)}~~,
\ee
and
\be
Hom_{c({\cal A})}(q, {\bf q})=\oplus_{\alpha\in A}
\oplus_{i\in J, j\in I_\alpha}{Hom_{\cal A}(b_i, a^{(\alpha)}_j)}~~,
\ee
which coincide with the spaces $Hom_{\cal B}(f,f)$ and $Hom_{\cal B}(q,f)$ 
computed above (see (\ref{ff}) and (\ref{qf})). We shall now show that the 
BRST operators on these spaces also agree. Indeed, on the space 
$Hom_{c({\cal A})}({\bf q}, {\bf q})$ we have the operator $Q_{{\bf q}, {\bf q}}$
given by:
{\footnotesize
\bea
Q_{{\bf q}, {\bf q}}u=\oplus_{\alpha,\beta \in A}\oplus_{i\in I_\alpha, j\in I_\beta}
{\left[
Q_{a^{(\alpha)}_i, a^{(\beta)}_j}u^{(\alpha\beta)}_{ij}+
\sum_{\gamma\in A, k\in I_\gamma}{{\bf q}_{\gamma k, \beta j}u^{(\alpha\gamma)}_{ik}}-
\sum_{\gamma\in A, k\in I_\gamma}{(-1)^{|u^{(\gamma\beta)}_{kj}|}u^{(\gamma\beta)}_{kj}
{\bf q}_{\alpha i, \gamma k}}
\right]}~~,\nn
\eea}\noindent where 
$u=\oplus_{\alpha,\beta\in A}{\oplus_{i\in I_\alpha, j\in I_\beta}
{u^{(\alpha\beta)}_{ij}}}$, with $u^{(\alpha\beta)}_{ij}\in 
Hom_{\cal A}(a^{(\alpha)}_i, a^{(\beta)}_j)$.
It is easy to see that this agrees with expression   (\ref{Q_{ff}}) for 
$Q_{ff}$ upon substituting our definition (\ref{q_new}) for ${\bf q}$.

On the space $Hom_{c({\cal A})}(q,{\bf q})$ we have the operator:
\be
Q_{q{\bf q}}u=\oplus_{\alpha \in A}\oplus_{i\in J, j\in I_\alpha}{\left
[Q_{b_i, a^{(\alpha)}_j}u^{(\alpha)}_{ij}+
\sum_{\beta \in A, k\in I_\beta}{{\bf q}_{\beta k, \alpha j}u^{(\beta)}_{ik}}-
\sum_{k\in J}{(-1)^{|u^{(\alpha)}_{kj}|}u^{(\alpha)}_{kj}q_{ik}}
\right]}~~,
\ee
where $u=\oplus_{\alpha \in A}\oplus_{i\in J, j\in I_\alpha}{u^{(\alpha)}_{ij}}$
with $u^{(\alpha)}_{ij}\in Hom_{\cal A}(b_i, a^{(\alpha)}_j)$.
This agrees with (\ref{Q_{qf}}) upon substituting the definition (\ref{q_new}) of 
${\bf q}$. 

A similar computation can be done for the morphism spaces $Hom_{\cal B}(f, q)$ 
and $Hom_{c({\cal A})}({\bf q}, q)$. This is almost identical 
with our computation of
$Hom_{\cal B}(q,f)$ and $Hom_{c({\cal A})}(q,{\bf q})$ above and 
leads once again to agreement. 

\

\noindent {\em 
Equivalence of ${\cal B}=c({\cal A})_f$ with a subcategory of $c({\cal A})$}

\

It is now obvious how to define the desired dG-equivalence. One considers the 
full subcategory ${\cal C}(f)$ of $c({\cal A})$ consisting of ${\bf q}$ and 
those degree one generalized complexes $q$ over ${\cal A}$ which are distinct from the 
components $q^{(\alpha)}$ of $f$. One maps each such complex to the associated object 
of ${\cal B}$ and the complex ${\bf q}$ to the object $f$ of ${\cal B}$. Moreover, 
one maps morphisms of ${\cal C}(f)$ into morphisms of ${\cal B}$ in the obvious manner.
The previous computations show that the resulting functor $F$ is a dG-functor, which is
clearly a dG-equivalence. Hence ${\cal B}=c({\cal A})_f$ is 
dG-equivalent with the full subcategory ${\cal C}(f)$
of $c({\cal A})$. Note that this equivalence depends on our initial choice of 
enumeration for the objects of $f$ (which allowed us to view $f$ as a generalized complex),
and we obtain such an equivalence for every choice of enumeration.  It is also 
easy to see that $F$ preserves the corresponding bilinear forms.
Since this holds for any connected pseudocomplex $f$ over $c({\cal A})$, we 
conclude that $c({\cal A})$ is quasiunitary.

\

\

\section{Conclusions}

We discussed the general framework of associative open string theory in the 
presence of D-branes, relating it to the mathematical theory of differential 
graded categories.
We showed that D-brane composite formation can be described systematically in 
this framework, through the 
general mechanism of condensation of boundary condition changing operators, and
showed how this description can be extracted through the standard 
procedure of shifting the string vacuum. 

We formulated a quasiunitarity constraint which encodes the condition that 
a string field theory in the presence of D-branes provides a self-consistent 
description of D-brane dynamics, in the sense that it is closed under formation
of D-brane composites. We showed that 
this constraint can be satisfied by considering an enlargement $c({\cal A})$ 
of a given D-brane category ${\cal A}$ , its {\em quasiunitary cover}, 
which can be constructed as a category of generalized complexes over ${\cal A}$.
Our results provide a general description of D-brane composite formation and 
represent a first step toward a better understanding of the structure of open 
string field theory in the presence of D-branes. This analysis is extremely 
general and can be applied to any associative string field theory. 

It is clear that such a structural analysis could form a good foundation for 
gaining a better understanding of various dynamical issues in D-brane 
physics. Perhaps the most immediate application concerns a better understanding of 
K-theory as a classification of D-brane charge. Indeed, it seems likely that 
the 
ultimate formulation of D-brane charge should be a version of K-theory for a 
differential graded category (this should be a generalization of Quillen's 
K-theory of exact categories \cite{Quillen_K, Srinvas})
\footnote{Quillen's categorical Q-construction 
gives the most general formulation of K-groups currently in wide use in 
mathematics. This requires an exact category, i.e. essentially a category 
with exact sequences. Exact categories seems unnatural from the point of view 
of this paper, since they are not produced in any direct fashion by the 
fundamental structure
of open string theory with D-branes (though applications to topological 
open B-models do produce such objects). }.
Since our approach gives a systematic treatment 
of D-brane composite formation (thus including the essential dynamics behind 
the original arguments for the relevance of K-theory in D-brane physics
\cite{Witten_K}) this 
offers the hope of a string-theoretic proof that a certain version of K-theory is 
indeed conserved by D-brane dynamics. Note that such an approach to K-theory 
charges would be extremely general, and in particular directly applicable 
to bosonic open string theory, for which the current understanding of K-theoretic 
charges is rather indirect \cite{K_bos}. Also note that, since such an approach 
would be based on the underlying category of the string field theory, it would 
take into account the effect of all massive string modes. This seems to be related 
to similar ideas proposed in \cite{Witten_K_review} and would represent an 
off-shell counterpart to the approach of \cite{Moore_top}.

On a more speculative note, let me mention that our 
procedure (namely inclusion of all D-brane 
composites in order to obtain a 
unitary description of D-brane dynamics) is in a certain 
sense related to `second quantization' of D-branes (enlarging the state 
space such as to allow for D-brane generation/annihilation processes)
\footnote{I thank C. ~Hofman for pointing this out.}. 
It is likely that there exists a connection between our procedure and M-theory, 
maybe through recent attempts to formulate the 
topological sector of M-theory  as a theory of a topological membrane \cite{JS}.

\

\acknowledgments{
The author wishes to thank J.~S.~Park for pointing out the relevance of 
string field theory to the homological mirror symmetry conjecture and 
for many interesting conversations. He also thanks Professor M.~Rocek for 
support and interest in his work and C.~Hofman for an illuminating remark. 
The author is supported by the Research Foundation under NSF grant 
PHY-9722101. }

\appendix

\section{Differential graded categories}

This appendix collects some basic facts on differential graded categories. 
Good sources for more information are \cite{Bondal_Kapranov,Keller_dg}.
We assume that the reader is familiar with the basic concepts of category theory, 
at the level of the first chapters in \cite{MacLane}.

{\bf Definition} A {\em graded category} is a category ${\cal A}$ 
whose morphism spaces are endowed with the structure of graded complex vector 
spaces, and such that all morphism compositions are bilinear maps of degree zero:
\be
|uv|=|u|+|v|~{\rm~for~}u\in Hom(b,c), v\in Hom(a,b)~~,
\ee 
where $|.|$ denotes the degree on morphisms.

{\bf Definition} A {\em differential graded } (dG) category is a graded 
category ${\cal A}$
whose morphism spaces $Hom(a,b)$ are graded differential complexes. 
Moreover, the linear applications 
$Hom(b,c)\otimes Hom(a,b)\rightarrow Hom(a,c)$ induced by morphisms compositions are 
are degree zero morphisms of graded complexes, where $Hom(b,c)\otimes Hom(a,b)$ carries 
the structure of total complex: 
\be
d(u\otimes v)=(du)\otimes v+(-1)^{|u|}u\otimes dv~~,
\ee
for $(u,v)\in Hom(b,c)\times Hom(a,b)$. 

Note that the differentials $d$ on $Hom(a,b)$ are taken to have degree $+1$. 
The condition on morphism compositions reads:
\be
\label{derivation}
d(uv)=(du)v +(-1)^{|u|}udv~~. 
\ee
That is, $d$ acts as a derivation of morphism composition.

One defines the notions of 
graded subcategory and dG subcategory in the obvious manner.

{\bf Definition} Given a dG category ${\cal A}$, 
its {\em cohomology category} is 
the graded category $H({\cal A})$ 
on the same objects as ${\cal A}$, and whose morphism 
spaces are given by the cohomology of the complexes $Hom_{\cal A}(a,b)$:
\be
Hom_{H({\cal A})}(a,b)=H_d^*(Hom_{\cal A}(a,b))
\ee
The morphism compositions of $H({\cal A})$ are induced by those of ${\cal A}$ (the induced 
compositions are well-defined due to condition (\ref{derivation})).

{\bf Definition} A (covariant) functor $F:{\cal A}\rightarrow {\cal B}$ between graded 
categories is a {\em graded functor} if the maps 
$F:Hom(a,b)\rightarrow Hom(F(a),F(b))$ are homogeneous of degree zero 
for any two objects $a, b$ of ${\cal A}$. It is a {\em graded equivalence} if it 
is graded and admits a graded inverse.

{\bf Definition} A functor $F:{\cal A}\rightarrow {\cal B}$ between two dG 
categories is a {\em dG functor} if: 

(1) It is a graded functor

(2) It commutes with the differentials on morphisms, i.e. :
\be
F(du)=dF(u)~~{\rm~for~all~objects~}~a,b~{\rm~of~}{\cal A}~~.
\ee

A dG functor $F$ descends to 
a well-defined graded functor $H(F):H({\cal A})\rightarrow 
H({\cal B})$ between the associated cohomology categories. Moreover, we 
have $H(FG)=H(F)H(G)$, i.e. taking cohomology is itself a functorial operation.

{\bf Definition} A dG functor $F:{\cal A}\rightarrow {\cal B}$ is 
a {\em dG equivalence} if it admits a dG-functor as an inverse.

\end{document}